\documentclass[aps,prd,reprint,nofootinbib,longbibliography]{revtex4-1}

\usepackage{amsmath}
\usepackage{mathtools}
\usepackage{graphicx}
\usepackage[normalem]{ulem}
\usepackage[com pat=1.1.0]{tikz-feynman}
\usepackage{tikz}
\usepackage{aas_macros}
\usetikzlibrary{external}
\usepackage{comment}
\usetikzlibrary{arrows}
\usepackage[colorlinks=true, allcolors=blue]{hyperref}

\begin{document}

\title{Correlation analysis of gravitational waves and neutrino signals to constrain neutrino flavor conversion in core-collapse supernova}

\author{Hiroki Nagakura}
\email{hiroki.nagakura@nao.ac.jp}
\affiliation{Division of Science, National Astronomical Observatory of Japan, 2-21-1 Osawa, Mitaka, Tokyo 181-8588, Japan}

\author{David Vartanyan}
\affiliation{Carnegie Observatories, 813 Santa Barbara St, Pasadena, CA 91101, USA; NASA Hubble Fellow}

\begin{abstract}
Recent multi-dimensional (multi-D) core-collapse supernova (CCSN) simulations characterize gravitational waves (GWs) and neutrino signals, offering insight into universal properties of CCSN independent of progenitor. Neutrino analysis in real observations, however, will be complicated due to the ambiguity of self-induced neutrino flavor conversion (NFC), which poses an obstacle to extracting detailed physical information. In this paper, we propose a novel approach to place a constraint on NFC from observed quantities of GWs and neutrinos based on correlation analysis from recent, detailed multi-D CCSN simulations. The proposed method can be used even in cases with low significance - or no detection of GWs. We also discuss how we can utilize electro-magnetic observations to complement the proposed method. Although our proposed method has uncertainties associated with CCSN modeling, the present result will serve as a base for more detailed studies. Reducing the systematic errors involved in CCSN models is a key to success in this multi-messenger analysis that needs to be done in collaboration with different theoretical groups.
\end{abstract}
\maketitle

\section{Introduction}\label{sec:intro}
The next galactic core-collapse supernova (CCSN) is a promising candidate bringing the first simultaneous detection of gravitational waves (GWs) and neutrinos. These signals not only directly probe the explosion mechanism of CCSNe, but also provide insight into microphysical properties such as warm nuclear matter and quantum features of neutrinos including flavor conversions. This great potential has motivated efforts to develop realistic theoretical models and high-fidelity detectors with various technologies.

There are many previous theoretical studies to understand characteristic features of GWs and neutrino signals (see reviews, e.g., \cite{2020arXiv201004356A} for GWs and \cite{2016NCimR..39....1M,2018JPhG...45d3002H,2019ARNPS..69..253M} for neutrinos and references therein). GW spectrogram analysis based on multi-dimensional (multi-D) simulations is one of major approaches to probe the interior physics of CCSNe \citep{2016ApJ...829L..14K,2017MNRAS.468.2032A,2019ApJ...876L...9R,2020PhRvD.102b3027M,2023arXiv230106515J}. This analysis can also be complemented by linear analysis (or asteroseismology) of the proto-neutron star (PNS) \cite{2015MNRAS.450..414F,2017PhRvD..96f3005S,2018MNRAS.474.5272T,2018ApJ...861...10M,2019MNRAS.482.3967T,2019PhRvL.123e1102T,2021PhRvD.103f3006B,2023PhRvD.107h3029B} or neutrino signal \citep{2016MNRAS.461.3296N,2017ApJ...851...62K,2020ApJ...898..139W}. These theoretical studies are important efforts to maximize the scientific gain from real observations. From the observational point of view, however, detections of GWs from CCSNe are technically more difficult than compact binary coalescence, due to the lack of definitive theoretical GW templates (see, e.g., \cite{2021PhRvD.104j2002S,2023arXiv230507688D}). This indicates that the GW signal may not be clear enough for analyzing detailed features including their time structures and frequency-dependent features (see also Fig.~5 in \cite{2020PhRvD.101h4002A} for detection efficiency as a function of distance during O1 and O2). While \citet{2017MNRAS.468.2032A} suggested that the current operating GW detectors have the ability to detect GW events only for $\lesssim 2$kpc CCSNe, other groups, e.g., \cite{2016ApJ...829L..14K,2020MNRAS.494.4665P,2023PhRvD.107d3008M}, suggested that the detection horizon is $\gtrsim 10$kpc. In pessimistic cases, we will obtain only the upper limit of the total radiated energy of GWs ($E_{\rm GW}$), posing a natural question: how we can utilize the upper limit of $E_{\rm GW}$ to extract physical information from real observations? This is what motivates the present study.

There is also physical motivation in the present study. Recent theoretical studies indicate that neutrino flavor conversion (NFC) in CCSNe is more complicated than the canonical picture with vacuum and matter effects (see, e.g., \cite{2000PhRvD..62c3007D}). Large flavor conversions can be triggered by neutrino self-interactions (see \cite{2016NuPhB.908..366C,2021ARNPS..71..165T,2022Univ....8...94C,2022arXiv220703561R} for recent reviews), and the associated flavor conversion instabilities ubiquitously occur in CCSNe \cite{2019ApJ...886..139N,2019ApJ...883...80S,2020PhRvD.101b3018D,2020PhRvR...2a2046M,2021PhRvD.104h3025N,2021PhRvD.103f3033A,2021PhRvD.103f3013C,2022ApJ...924..109H,2023PhRvD.107h3016X,2023PhRvL.130u1401N}. Since these neutrino flavor conversions hinge on multiple factors, e.g., the neutrino energy spectrum, angular distributions, and neutrino-matter interactions, it is hard to determine a priori the mixing degree of neutrinos via theoretical models. This indicates that the survival probability of neutrinos needs to be treated as an unknown variable in real observations. NFC is, hence, a key uncertainty to extract physical information from the neutrino signal. It is also worth to note that, although neutral-current channels have the ability to provide NFC-independent features of neutrino signal, they are lower statistics than major detection channels mostly sensitive to only electron-type neutrinos and their antipartners. Our proposed method will be complementary to these analyses.

\begin{figure*}
\begin{minipage}{0.45\textwidth}
    \includegraphics[width=\linewidth]{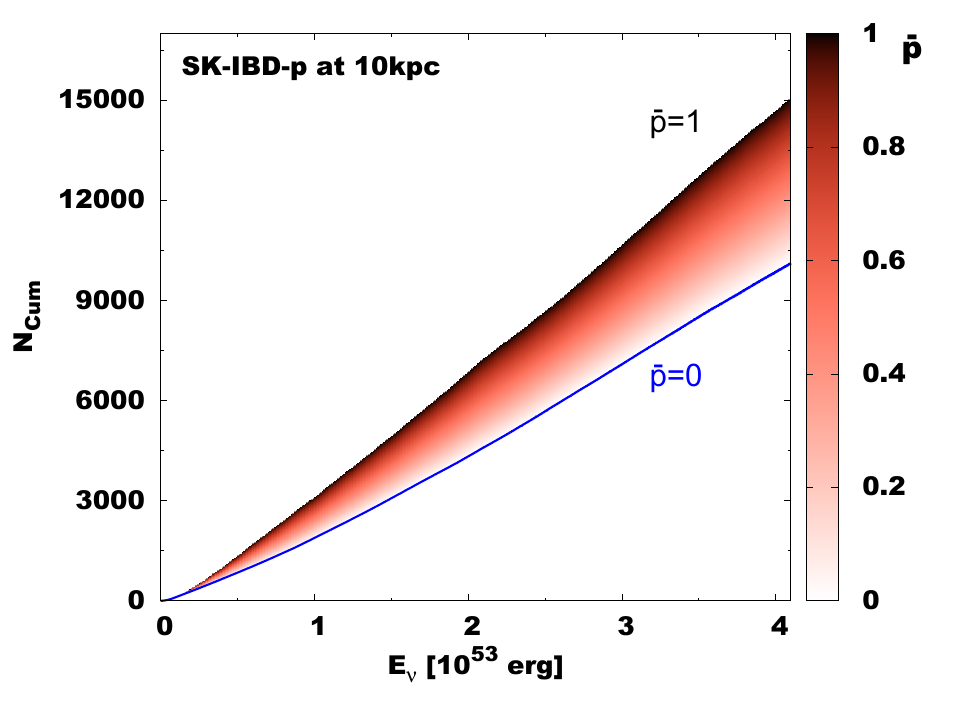}
\end{minipage}
\begin{minipage}{0.45\textwidth}
    \includegraphics[width=\linewidth]{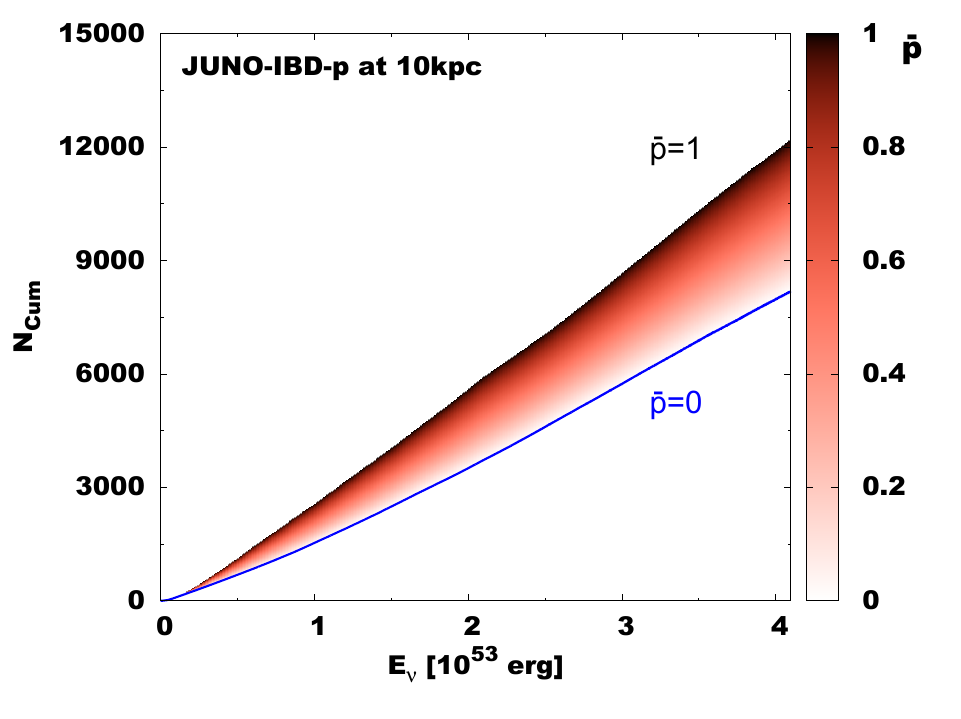}
\end{minipage}
    \caption{$N_{\rm Cum}$ versus $E_{\nu}$, which is less dependent on progenitors \cite{2021MNRAS.500..696N,2021MNRAS.506.1462N} (but see some caveats in Sec.~\ref{subsec:lim}). The color represents the survival probability of electron-type anti-neutrinos ($\bar{p}$). We highlight the case with $\bar{p}=0$ as a blue solid line. We assume a CCSN source distance of $10$kpc. White-colored regions corresponds to the forbidden area from the correlation. We focus
on a detection channel of IBD-p. Left: SK. Right: JUNO. The detector volume is assumed to be $32.5$ktons and $20$ktons, respectively.
}
    \label{graph_2Dcolormap_Enu_vs_Ncum}
\end{figure*}

In this paper, we present a correlation analysis of GWs and neutrino signals to PNS structure by using results of most recent multi-D CCSN simulations, aiming to place a constraint on NFC. We make a statement that GW signal can break the degeneracy between NFC and neutrino detection counts. The proposed method can be used even in cases with low significance $-$ or no detection of GWs, that is of great use in the data analysis. We note that CCSN models, and hence our results, depend on uncertainties in input physics such as the nuclear equation-of-state (EOS) (see also \cite{2021ApJ...923..201E}) and neutrino reaction rates. Improved experimental and theoretical constraints and broader collaboration between CCSNe modeling groups are necessary to resolve these uncertainties. Our aim with this study is also to help guide the collaboration strategy among GWs/neutrinos observations and theoretical studies.

\begin{figure*}
\begin{minipage}{0.42\textwidth}
    \includegraphics[width=\linewidth]{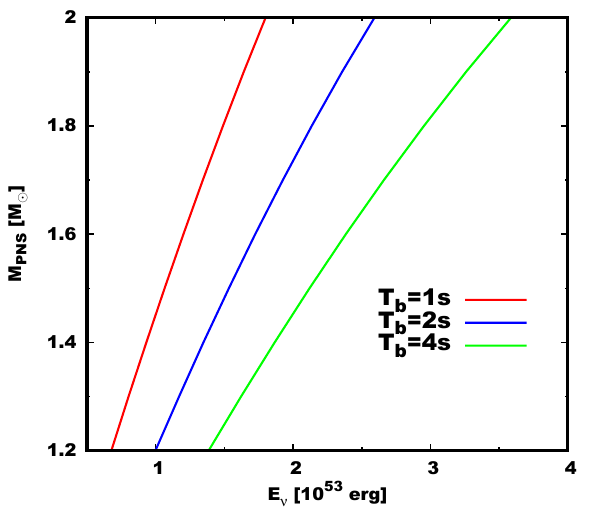}
\end{minipage}
\begin{minipage}{0.42\textwidth}
    \includegraphics[width=\linewidth]{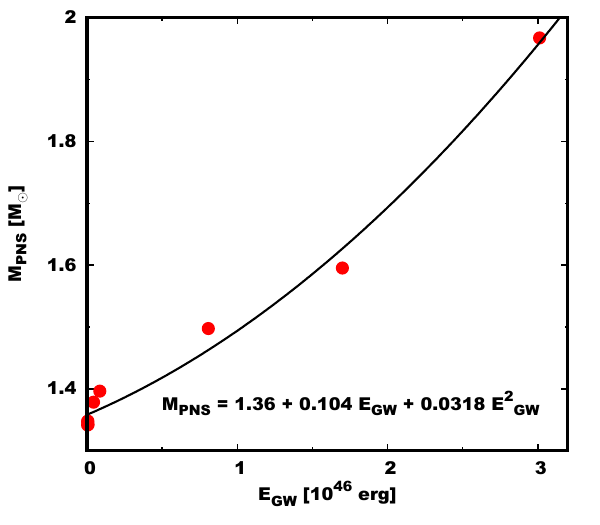}
\end{minipage}
    \caption{Left: Plots of flavor- and time-integrated total neutrino energy ($E_{\nu}$) versus PNS mass ($M_{\rm PNS}$). Different color indicates different post-bounce time ($T_{\rm b}$) at which the correlation between $M_{\rm PNS}$ and $E_{\nu}$ is displayed. Right: Plots of radiated energy of GW ($E_{\rm GW}$) versus PNS mass ($M_{\rm PNS}$) for CCSN models in \cite{2023arXiv230207092V} (red filled circles). We fit them quadratically (black solid line).
}
    \label{graph_EGW_vs_MPNS}
\end{figure*}

\section{Correlation study from CCSN models}\label{sec:CorStudy}
\subsection{CCSN models}\label{subsec:CCSNmodel}
At present, performing multi-D numerical simulations is the only possible way to quantify the GWs and neutrino signals. During the last few years, neutrino-radiation-hydrodynamic code, F{\sc{ornax}} \cite{2019ApJS..241....7S} has been used extensively to carry out a series of numerical simulations of CCSNe. We quantified these observable signals with explosion models across a wide range of progenitor masses \cite{2020MNRAS.491.2715B,2021Natur.589...29B}. We solve neutrino transport by a multi-energy (12 energy group) and multi-species (3 species) two-moment approximation with M1 closure relation \cite{2011JQSRT.112.1323V}. For all models, SFHo equation-of-state (EOS) \cite{2013ApJ...774...17S}, which is consistent with nuclear experiments and astrophysical constraints, is used. Reaction rates of neutrino-matter interactions are taken from \cite{2006NuPhA.777..356B} but with some extensions such as many body corrections in their axial vector part \cite{2017PhRvC..95b5801H} and electron-capture by heavy nuclei \cite{2010NuPhA.848..454J}. For more detailed information on F{\sc{ornax}}, we refer readers to a series of our papers, e.g., \cite{2019ApJS..241....7S}.

\subsection{Neutrino signal}\label{subsec:neutrinos}
In \cite{2021MNRAS.500..696N,2021MNRAS.506.1462N,2023arXiv230708735V}, we presented an in-depth analysis of neutrino signal from CCSNe and found that there is a robust correlation between the total neutrino energy (TONE or $E_{\nu}$) and the cumulative number of events ($N_{\rm Cum}$) in each detector. This correlation, however, depends on NFC model; for instance, $N_{\rm Cum}$ for a reaction channel by inverse-beta decay on protons (IBD-p), which corresponds to the major reaction for supernova neutrinos in Super-Kamiokande (SK) \cite{2016APh....81...39A} and Jiangmen Underground Neutrino Observatory (JUNO) \cite{2016JPhG...43c0401A}, widely varies depending on the survival probability of neutrinos.
To see the variation more quantitatively, we compute $N_{\rm Cum}$ by varying the survival probability of electron-type anti-neutrinos ($\bar{p}$) from 0 (maximum flavor conversion) to 1 (no flavor conversion) for the 20 M$_{\odot}$ progenitor in \cite{2021Natur.589...29B,2021MNRAS.506.1462N}, and the result is summarized in Fig.~\ref{graph_2Dcolormap_Enu_vs_Ncum}. As shown in this figure, determining $E_{\nu}$ from $N_{\rm Cum}$ has an uncertainty of $\sim 40 \%$ due to $\bar{p}$.

In our previous study \cite{2022MNRAS.512.2806N}, we also found that PNS mass ($M_{\rm PNS}$) has a strong correlation with $E_{\nu}$. In the left panel of Fig.~\ref{graph_EGW_vs_MPNS}, we show $M_{\rm PNS}$ as a function of $E_{\nu}$ by using the result in \cite{2022MNRAS.512.2806N}. We note that $E_{\nu}$ corresponds to the total emitted neutrino energy up to a certain post-bounce time ($T_{\rm b}$ which is measures from the time of core bounce), and we display three cases with $T_{\rm b}=1, 2,$ and $4$s in the figure. To determine $T_{\rm b}$, we need to identify the core-bounce time in real observations, which is expected to be determined within the uncertainty of a few tens of milliseconds \cite{2009PhRvD..80h7301H}. Since the uncertainty is much smaller than $T_{\rm b}$ which we consider ($> 1$s), it does not affect our correlation study. One thing we need to mention here is, however, that the neutrino survival probability needs to be a priori set if we determine $M_{\rm PNS}$ from $N_{\rm Cum}$. This motivates the use of GW signal to remove the ambiguity of neutrino oscillation.

\subsection{Gravitational waves}\label{subsec:GWs}

Let us turn our attention to GWs. The characteristic property of GWs in F{\sc{ornax}} CCSN models have been studied in \cite{2018ApJ...861...10M,2019ApJ...876L...9R,2019MNRAS.489.2227V,2020ApJ...901..108V}, and very recently \citet{2023arXiv230207092V} carried out a systematic study with long-term 3D simulations ($> 1$s) and we quantify the total emitted energy of GWs ($E_{\rm GW}$). Although the high computational cost still limits the number of models, we find a robust correlation between $E_{\rm GW}$ and the compactness of the progenitor for explosion models. Observationally, this is useful, since $E_{\rm GW}$ may be the most easily constrained in real observations even for cases with no detections \cite{2020PhRvD.101h4002A}. We note that $E_{\rm GW}$ is dominated by aspherical matter motions in the frequency range of $\gtrsim 100$ Hz, whereas the low frequency components including GW emission by anisotropic neutrino emission has a negligible contribution \cite{1997A&A...317..140M,2020arXiv201004356A,2020ApJ...901..108V}.

Let us describe the rationale behind the correlation between $E_{\rm GW}$ and the compactness of presupernova progenitor. The progenitor with the higher compactness core, in general, has higher mass accretion onto PNS in the post-bounce phase (see also \cite{2022MNRAS.517..543W}), that also leads to heavier $M_{\rm PNS}$. Strong turbulent energy fluxes are accompanied by the large mass accretion onto PNS for explosion models, which is the major driving force emitting GWs. Here, we should make an important remark. The turbulence in post-shock region tends to be weak for non-exploding (or black hole formation) cases \cite{2023arXiv230207092V}, since the accretion is more spherical and the post-shock accretion flow has higher temperature (i.e., low Mach number) than those in explosion models. This indicates that the correlation between $E_{\rm GW}$ and the compactness disappears in non-exploding models. For this reason, we adopt only explosion models in this correlation study. Although it is a limitation of the present work, the failure of explosion seems to be perhaps rarer than ordinary CCSNe \cite{2012ApJ...757...69U,2021Natur.589...29B}; hence, our proposed method will be applied in the majority of the death of massive stars.

\begin{figure*}
\begin{minipage}{0.95\textwidth} 
    \includegraphics[width=\linewidth]{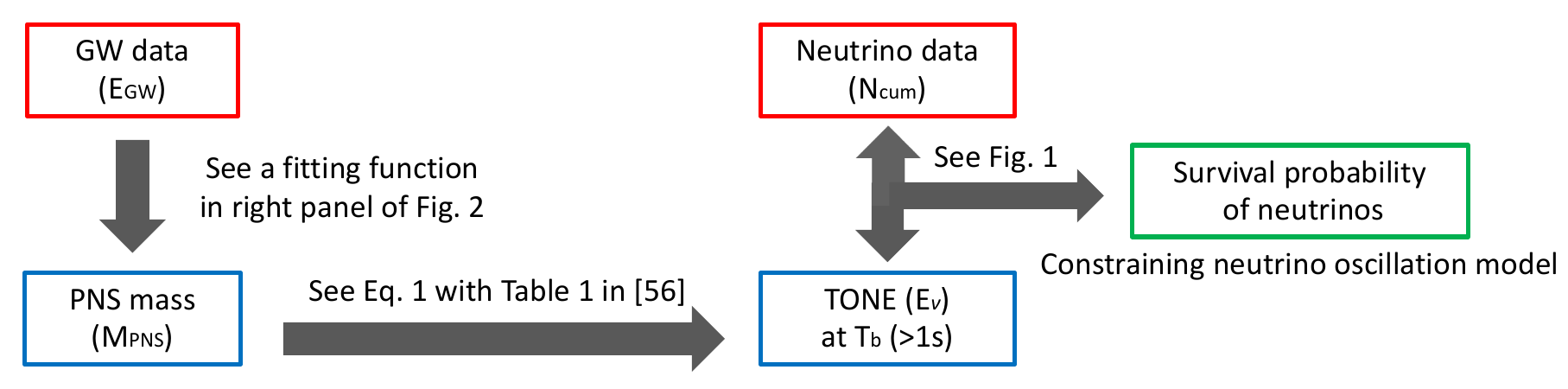}
\end{minipage}
    \caption{Flowchart of our proposed analysis. Red, blue, and green borders of each square distinguish observed quantities, theories, and output, respectively.
}
    \label{Schematic}
\end{figure*}

In the right panel of Fig.~\ref{graph_EGW_vs_MPNS}, we plot $M_{\rm PNS}$ (in the unit of solar mass, $M_{\odot}$) as a function of radiated GW energy ($E_{\rm GW}$ in the unit of $10^{46} {\rm erg}$) for 3D explosion models in \cite{2023arXiv230207092V}. We note that GW strain is estimated by using the quadrupole approximation \cite{1990ApJ...351..588F}. The positive correlation can be clearly seen, and we show the quadratic fit as a black solid line in this figure. We note that the minimum mass of $M_{\rm PNS}$ obtained from the fitting function, $1.36 M_{\odot}$, is not physical but rather an artifact due to the accuracy of polynominal fitting. The actual minimum PNS can be lower. We also quantify the coefficient of determination and standard deviation for the fitting function, which are $0.988$ and $0.018$, respectively. The latter is estimated based on a normalized error defined as $(M_{\rm PNS(d)} - M_{\rm PNS(f)})/M_{\rm PNS(f)}$, where $M_{\rm PNS(d)}$ and $M_{\rm PNS(f)}$ denote PNS mass at data point and that estimated by the fitting function, respectively.

\subsection{Demonstration}\label{subsec:demo}

Below we describe how to place a constraint on $\bar{p}$ by using these three progenitor-independent correlations. We provide a flowchart of our proposed method in Fig.~\ref{Schematic}. For readers seeking more detailed understandings of our method, necessary references at each procedure are also described. As the first step, we need to set $T_{\rm b}$. According to \cite{2023arXiv230207092V}, $E_{\rm GW}$ is mostly saturated up to $T_{\rm b} \sim 2$s, meanwhile the correlation of neutrino signal which we discussed in \cite{2021MNRAS.506.1462N,2022MNRAS.512.2806N} is guaranteed up to $T_{\rm b} \sim 4$s; hence it should be set in the range of $2 {\rm s} \lesssim T_{\rm b} \lesssim 4$s. Next, we estimate $M_{\rm PNS}$ from $E_{\rm GW}$ (see the right panel in Fig.~\ref{graph_EGW_vs_MPNS}), and then $E_{\nu}$ can be obtained from the correlation to $M_{\rm PNS}$ for given $T_{\rm b}$ (see the left panel in Fig.~\ref{graph_EGW_vs_MPNS}). $E_{\nu}$ provides the expected range of $N_{\rm Cum}$ depending on survival probability of neutrinos (see Fig.~\ref{graph_2Dcolormap_Enu_vs_Ncum}), and we can determine $\bar{p}$ by using the observed $N_{\rm Cum}$.

\begin{figure*}
\begin{minipage}{0.45\textwidth}
    \includegraphics[width=\linewidth]{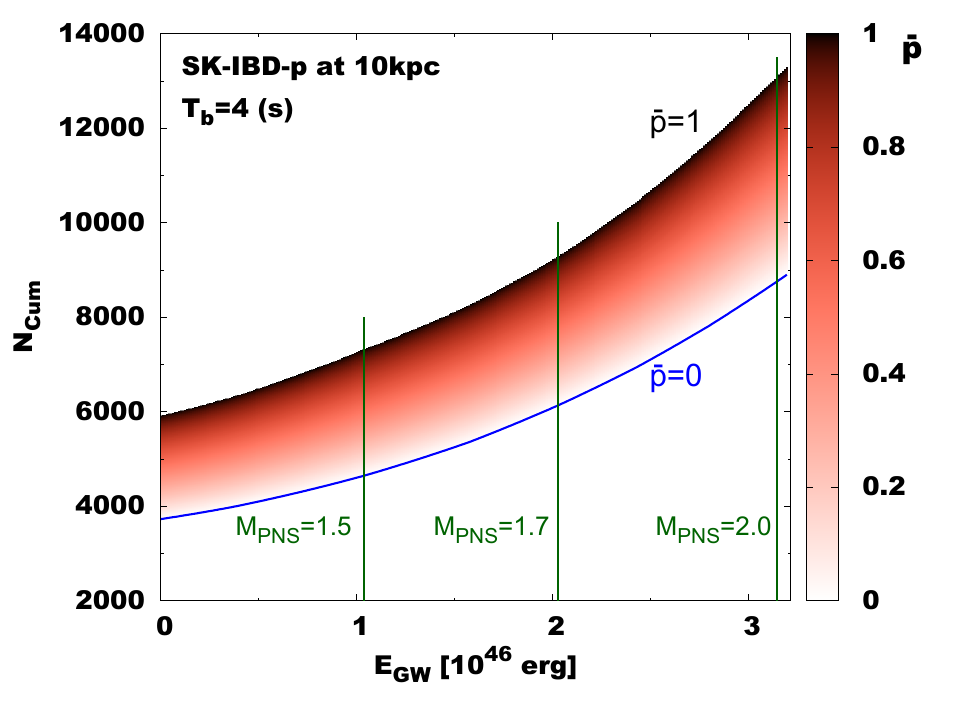}
\end{minipage}
\begin{minipage}{0.45\textwidth}
    \includegraphics[width=\linewidth]{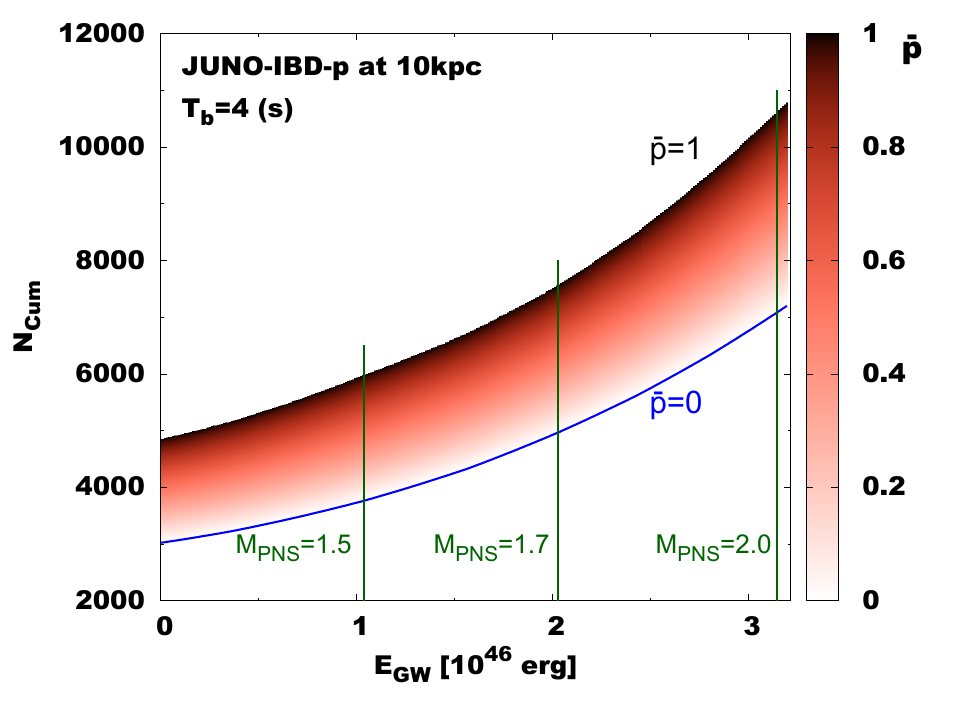}
\end{minipage}
    \caption{$N_{\rm Cum}$ versus $E_{\rm GW}$. Color map and line types are the same as those used in Fig.~\ref{graph_2Dcolormap_Enu_vs_Ncum}. The vertical green lines at each panel highlight $M_{\rm PNS}=1.5, 1.7,$ and $2.0 M_{\odot}$, which is one-to-one relation with $E_{\rm GW}$ (see the left panel of Fig.~\ref{graph_EGW_vs_MPNS}). Left and right panels show the case with SK and JUNO, respectively for $T_{\rm b}=4$s.
}
    \label{graph_2Dcolormap_EGW_vs_Ncum}
\end{figure*}

Fig.~\ref{graph_2Dcolormap_EGW_vs_Ncum} depicts the summary of our proposed method. The color map displays $\bar{p}$ as a function of $E_{\rm GW}$ and $N_{\rm Cum}$ for a representative case of $T_{\rm b}=4$s. To guide the eye, the correspondence between $M_{\rm PNS}$ and $E_{\rm GW}$ can be seen as green vertical lines for three values of $M_{\rm PNS}$: $1.5, 1.7,$ and $2.0 M_{\odot}$. This clearly illustrates that $\bar{p}$ can be obtained from $T_{\rm b}$ and two observed quantities, $E_{\rm GW}$ and $N_{\rm Cum}$. We note that, in real observations, $E_{\rm GW}$ will be determined with errors, which depends on the signal-to-noise ratio (S/N). If $E_{\rm GW}$ is determined within the error of a few tens of percent, $\bar{p}$ can be constrained very well.

Finally, let us consider how we can use our proposed method in cases with no detection of GWs. We first need to estimate the distance to the CCSN source, which is expected to be given by neutrino and electro-magnetic signals (see also Sec.~\ref{subsec:lim} for more details). The obtained distance can be used to estimate the upper limit of $E_{\rm GW}$ (see, e.g., \cite{2020PhRvD.101h4002A}), that also provides the upper limit of $M_{\rm PNS}$ (see right panel of Fig.~\ref{graph_EGW_vs_MPNS}). On the other hand, the observed $N_{\rm Cum}$ gives another upper limit of $M_{\rm PNS}$ only through the correlation between $M_{\rm PNS}$ and neutrino signal. Our proposed method becomes very meaningful if the upper limit of $M_{\rm PNS}$ constrained by GW is smaller than that constrained by only the neutrino signal. In this case, $N_{\rm Cum}$ and the upper limit of $E_{GW}$ offers the lower limit of $\bar{p}$, i.e., narrowing down the possible $\bar{p}$ candidate. This is very informative to determine NFC in CCSN core.

\subsection{Limitations}\label{subsec:lim}
We add some important remarks in the present study. First,
both detector- and Poisson noises in SK and JUNO do not compromise the estimation of $E_{\nu}$ from $N_{\rm Cum}$. Although the detector noise is a critical issue to retrieve energy spectrum of neutrinos from observed data \cite{2021MNRAS.500..319N}, we only need energy-integrated number of events in our method, which is not affected by the detector noise. The Poisson noise of $N_{\rm Cum}$ is also very small (less than a few percent), since the total number of event counts is more than a few thousands for Galactic CCSNe. It should be mentioned, however, that neutrino detections by other channels, e.g., electron-scatterings, need to be separated for the precise determination of IBD-p events. The gadolinium (Gd) doping in water Cherenkov detector improves the neutron-tagging efficiency, which increases the accuracy of separation between IBD-p and electron-scattering events \cite{2004PhRvL..93q1101B}. Gd has already been loaded in SK, and SK-VI (the first SK-Gd project) operated from August 2020 to June 2022 with $5.4$ tons of gadolinium ($0.01 \%$ mass concentration) \cite{2023arXiv230505135H} with a neutron capture efficiency of $\sim 50 \%$. At the moment, SK-Gd is operating with $0.033 \%$ mass concentration of Gd ($\sim 75 \%$ neutron capture efficiency), and the SK-Gd project plans to increase the concentration up to $\sim 0.1 \%$ in future, potentially offering $\sim 90 \%$ neutron capture efficiency \cite{2023PTEP.2023a3H01H}. Angular distributions of neutrino events also offer another means to distinguish IBD-p and electron-scattering. We can, hence, expect that IBD-p events will be identified well, although the quantitative discussion should be made in real observations\footnote{We note that, if we quantify the correlation between $E_{\nu}$ and $N_{\rm Cum}$ including electron-scattering events, we do not have to carry out the separation of the two reaction channels. On the other hand, we need to make sure whether there is a robust correlation between $E_{\nu}$ and $N_{\rm Cum}$. The detailed study is postponed to future work.}.

Second, there is some degree of progenitor dependence in the correlation between $E_{\nu}$ and $N_{\rm Cum}$, which is $\lesssim 10 \%$ for IBD-p events in SK and JUNO (see Fig.14 in \cite{2021MNRAS.506.1462N})\footnote{We note that the correlation becomes weaker if we include non-explosion models. As we have already mentioned, however, our proposed method can be used only for explosion models, since the correlation between $E_{\rm GW}$ and $M_{\rm PNS}$ almost disappers for non-exploding models.}. The progenitor dependence tends to be remarkable in the case with no flavor conversion (i.e., $\bar{p}=1$), whereas it monotonically decreases with $\bar{p} \to 0$. This is attributed to the two reasons. The first one is that IBD-p becomes sensitive to heavy-leptonic neutrinos at the CCSN source, which dominates the irradiated energy of neutrinos. This enhances the correlation between $E_{\nu}$ and $N_{\rm Cum}$. The other reason is that neutrino luminosity of heavy-leptonic neutrinos is dominated by the diffusion component, while the accretion component plays a non-negligible role in electron-type neutrinos and antineutrinos. The accretion component depends on the density profile of progenitor, indicating that the progenitor dependence tends to be stronger. Since the case with no flavor conversion, $N_{\rm Cum}$ is determined only by electron-type neutrinos at the source, leading to relatively larger progenitor dependence. This consideration suggests that, if $\bar{p}=1$ is obtained in real data analysis, we need to keep in mind the progenitor-dependent uncertainty.

Third, $N_{\rm Cum}$ in general depends on the observer direction, and the angular variation of the time-integrated event rate would be more than $10 \%$ for some CCSNe \cite{2019MNRAS.489.2227V,2021MNRAS.500..696N}. This indicates that the uncertainty of observer location potentially becomes a major systematic error in our proposed method. It is worth to note, however, that the asymmetry of neutrino emission can be estimated if low frequency (below $\sim 10$ Hz) GWs are detected. This is because the so-called memory effect of neutrino emision seems to dominate GWs in these frequency range \cite{1987Natur.327..123B,1997A&A...317..140M,2007ApJ...655..406K,2009ApJ...697L.133K,2012A&A...537A..63M,2020ApJ...901..108V,2022PhRvD.105l3028F}. This indicates that the multi-band GW analysis offers valuable information for our proposed method. In fact, it is impossible to constrain the asymmetry of neutrino emission only by neutrino observations.

Fourth, precise estimations of $E_{\nu}$ and $E_{\rm GW}$ require accurate determinations of distance to the CCSN source. The distance may be constrained by the neutronization burst within the error of $\sim 5 \%$ \cite{2005PhRvD..71f3003K}\footnote{It should be noted that the authors in \cite{2005PhRvD..71f3003K} assumed a megaton water Cherenkov detector in this estimation. This indicates that the error in real observations would be larger than their estimation even for Hyper-Kamiokande (its fiducial volume for supernova neutrino analysis is planned to be $\sim 220$ktons).}. Electro-magnetic wave (EM) observations can also provide another independent measurement; for instance, the angular diameter distance can be estimated from the angular size of the ejecta and its spatial size. The former is given by observations, and the latter is estimated by two quantities: the time since the explosion and the expansion veloity of the ejecta deduced from theoretical considerations. We also note that the pre-explosion image of CCSN progenitor may be available in the catalog of nearby red supergiants. This observation also offers the distance information independently from others. Although the precision of the measurement hinges on the CCSN event, the cross-check among these independent measuremets will reduce the distance uncertainty.

Fifth, one has to keep in mind another caveat in the present study. $\bar{p}$ is assumed to be constant in time when we quantify the correlation between $E_{\nu}$ from $N_{\rm Cum}$. However, $\bar{p}$ is in general time-dependent. This indicates that our proposed method has the ability to constrain only the time-averaged $\bar{p}$, and that systematic errors may become large if $\bar{p}$ varies with time considerably. The maximum error is quantified in principle by changing $\bar{p}$ as a function of time in our correlation study. We leave this problem for the future.

Sixth, we summarize potential systematic errors inherent in our CCSN models. The sensitivity to input physics (EOS and neutrino-matter interactions) to the correlation remains uncertain; indeed, we focused only the SFHo EOS and neutrino-matter interactions in our CCSN models, that would affect the correlations quantitatively. As another caveat, rotation may be an important contributor to GW and neutrino signals. However, our correlation study is made for non-rotating progenitors. We also note that CCSN models may have endemic unknown systematic errors due to complex numerical implementations of input physics.

Finally, we discuss how these systematic uncertainties and errors in observed quantities affect constraints on $\bar{p}$, which provides the range of application for our proposed method. As displayed in Fig.~\ref{graph_2Dcolormap_EGW_vs_Ncum}, $\sim 30 \%$ error in $N_{\rm Cum}$ overwhelms the variation by $\bar{p}$. This indicates that other potential uncertainties (e.g., distance error, angular variations, progenitor dependences, and systematic errors in CCSN models) become a critical problem in our proposed method, if they lead to the same amount of uncertainty in $N_{\rm Cum}$. For similar reasons, the error in $E_{\rm GW}$ (and an associated uncertainty in the correlation between $E_{\rm GW}$ and $M_{\rm PNS}$) needs to be within $\sim 40 \%$ to place a constraint on $\bar{p}$. As already pointed out, however, the upper limit of $E_{\rm GW}$ is still meaningful even if GWs are not observed.

Last but not least, it is very important to carry out a similar correlation study based on models simulated by different CCSN groups so as to reduce the systematic uncertainties in our proposed method. The survival probability of neutrinos and PNS mass will be estimated independently in each group, and then they will be compared among different groups. Such a collaborative work is indispensable to quantify the theoretical uncertainty, and it also increases statistics in the correlation analysis. We leave all these detailed investigations and fruitful collaborations as future work.

\section{Summary}\label{sec:summary}
In this paper, we propose a new approach of GW-neutrino joint analysis for CCSNe, based on the correlation study of most recent multi-D CCSN explosion models. Our method is designed so as to determine the survival probability of neutrinos and PNS mass, which are both key ingredients to characterize CCSN dynamics. One of the useful features in the proposed method is that we can apply it even in cases with no detection of GWs. This is a great advantage compared to other proposed GW-neutrino analysis, for instance GW spectrogram analysis \cite{2016ApJ...829L..14K,2017MNRAS.468.2032A,2019ApJ...876L...9R,2020PhRvD.102b3027M,2023arXiv230106515J,2015MNRAS.450..414F,2017PhRvD..96f3005S,2018MNRAS.474.5272T,2018ApJ...861...10M,2019MNRAS.482.3967T,2019PhRvL.123e1102T,2021PhRvD.103f3006B,2023PhRvD.107h3029B}, that commonly requires GW data with high S/N.

Finally, we put a comment on how EM signal can complement the proposed method. In recent CCSN simulations, a positive correlation between PNS mass and explosion energy of CCSNe can be seen, albeit somewhat weaker correlations than GWs and neutrinos. According to \cite{2022MNRAS.517..543W}, the explosion energy can reach $> 10^{51} {\rm erg}$ only if $M_{\rm PNS}$ is larger than $\sim 1.7 M_{\odot}$. This exhibits that the EM observations offer an independent constraint on $M_{\rm PNS}$, and they can be used as a consistency check to that obtained from the present method. As such, there is surely room for improvements, but this study offers a feasible way for multi-messenger analyses with numerical CCSN models.

\section{DATA AVAILABILITY}
The data underlying this article will be shared on reasonable request to the corresponding author.

\section{Acknowledgments}
We are grateful for ongoing contributions to the effort of CCSN simulation projects by Adam Burrows, David Radice, Josh Dolence, Aaron Skinner, Matthew Coleman, Chris White, and Tianshu Wang. HN is supported by Grant-inAid for Scientific Research (23K03468). DV acknowledges support from the NASA Hubble Fellowship Program grant HST-HF2-51520.

\bibliography{bibfile}

\begin{thebibliography}{73}%
\makeatletter
\providecommand \@ifxundefined [1]{%
 \@ifx{#1\undefined}
}%
\providecommand \@ifnum [1]{%
 \ifnum #1\expandafter \@firstoftwo
 \else \expandafter \@secondoftwo
 \fi
}%
\providecommand \@ifx [1]{%
 \ifx #1\expandafter \@firstoftwo
 \else \expandafter \@secondoftwo
 \fi
}%
\providecommand \natexlab [1]{#1}%
\providecommand \enquote  [1]{``#1''}%
\providecommand \bibnamefont  [1]{#1}%
\providecommand \bibfnamefont [1]{#1}%
\providecommand \citenamefont [1]{#1}%
\providecommand \href@noop [0]{\@secondoftwo}%
\providecommand \href [0]{\begingroup \@sanitize@url \@href}%
\providecommand \@href[1]{\@@startlink{#1}\@@href}%
\providecommand \@@href[1]{\endgroup#1\@@endlink}%
\providecommand \@sanitize@url [0]{\catcode `\\12\catcode `\$12\catcode
  `\&12\catcode `\#12\catcode `\^12\catcode `\_12\catcode `\%12\relax}%
\providecommand \@@startlink[1]{}%
\providecommand \@@endlink[0]{}%
\providecommand \url  [0]{\begingroup\@sanitize@url \@url }%
\providecommand \@url [1]{\endgroup\@href {#1}{\urlprefix }}%
\providecommand \urlprefix  [0]{URL }%
\providecommand \Eprint [0]{\href }%
\providecommand \doibase [0]{http://dx.doi.org/}%
\providecommand \selectlanguage [0]{\@gobble}%
\providecommand \bibinfo  [0]{\@secondoftwo}%
\providecommand \bibfield  [0]{\@secondoftwo}%
\providecommand \translation [1]{[#1]}%
\providecommand \BibitemOpen [0]{}%
\providecommand \bibitemStop [0]{}%
\providecommand \bibitemNoStop [0]{.\EOS\space}%
\providecommand \EOS [0]{\spacefactor3000\relax}%
\providecommand \BibitemShut  [1]{\csname bibitem#1\endcsname}%
\let\auto@bib@innerbib\@empty
\bibitem [{\citenamefont {{Abdikamalov}}\ \emph {et~al.}(2020)\citenamefont
  {{Abdikamalov}}, \citenamefont {{Pagliaroli}},\ and\ \citenamefont
  {{Radice}}}]{2020arXiv201004356A}%
  \BibitemOpen
  \bibfield  {author} {\bibinfo {author} {\bibfnamefont {Ernazar}\ \bibnamefont
  {{Abdikamalov}}}, \bibinfo {author} {\bibfnamefont {Giulia}\ \bibnamefont
  {{Pagliaroli}}}, \ and\ \bibinfo {author} {\bibfnamefont {David}\
  \bibnamefont {{Radice}}},\ }\bibfield  {title} {\enquote {\bibinfo {title}
  {{Gravitational Waves from Core-Collapse Supernovae}},}\ }\href {\doibase
  10.48550/arXiv.2010.04356} {\bibfield  {journal} {\bibinfo  {journal} {arXiv
  e-prints}\ ,\ \bibinfo {eid} {arXiv:2010.04356}} (\bibinfo {year} {2020})},\
  \Eprint {http://arxiv.org/abs/2010.04356} {arXiv:2010.04356 [astro-ph.SR]}
  \BibitemShut {NoStop}%
\bibitem [{\citenamefont {{Mirizzi}}\ \emph {et~al.}(2016)\citenamefont
  {{Mirizzi}}, \citenamefont {{Tamborra}}, \citenamefont {{Janka}},
  \citenamefont {{Saviano}}, \citenamefont {{Scholberg}}, \citenamefont
  {{Bollig}}, \citenamefont {{H{\"u}depohl}},\ and\ \citenamefont
  {{Chakraborty}}}]{2016NCimR..39....1M}%
  \BibitemOpen
  \bibfield  {author} {\bibinfo {author} {\bibfnamefont {A.}~\bibnamefont
  {{Mirizzi}}}, \bibinfo {author} {\bibfnamefont {I.}~\bibnamefont
  {{Tamborra}}}, \bibinfo {author} {\bibfnamefont {H.~Th.}\ \bibnamefont
  {{Janka}}}, \bibinfo {author} {\bibfnamefont {N.}~\bibnamefont {{Saviano}}},
  \bibinfo {author} {\bibfnamefont {K.}~\bibnamefont {{Scholberg}}}, \bibinfo
  {author} {\bibfnamefont {R.}~\bibnamefont {{Bollig}}}, \bibinfo {author}
  {\bibfnamefont {L.}~\bibnamefont {{H{\"u}depohl}}}, \ and\ \bibinfo {author}
  {\bibfnamefont {S.}~\bibnamefont {{Chakraborty}}},\ }\bibfield  {title}
  {\enquote {\bibinfo {title} {{Supernova neutrinos: production, oscillations
  and detection}},}\ }\href {\doibase 10.1393/ncr/i2016-10120-8} {\bibfield
  {journal} {\bibinfo  {journal} {Nuovo Cimento Rivista Serie}\ }\textbf
  {\bibinfo {volume} {39}},\ \bibinfo {pages} {1--112} (\bibinfo {year}
  {2016})},\ \Eprint {http://arxiv.org/abs/1508.00785} {arXiv:1508.00785
  [astro-ph.HE]} \BibitemShut {NoStop}%
\bibitem [{\citenamefont {{Horiuchi}}\ and\ \citenamefont
  {{Kneller}}(2018)}]{2018JPhG...45d3002H}%
  \BibitemOpen
  \bibfield  {author} {\bibinfo {author} {\bibfnamefont {Shunsaku}\
  \bibnamefont {{Horiuchi}}}\ and\ \bibinfo {author} {\bibfnamefont {James~P.}\
  \bibnamefont {{Kneller}}},\ }\bibfield  {title} {\enquote {\bibinfo {title}
  {{What can be learned from a future supernova neutrino detection?}}}\ }\href
  {\doibase 10.1088/1361-6471/aaa90a} {\bibfield  {journal} {\bibinfo
  {journal} {Journal of Physics G Nuclear Physics}\ }\textbf {\bibinfo {volume}
  {45}},\ \bibinfo {pages} {043002} (\bibinfo {year} {2018})},\ \Eprint
  {http://arxiv.org/abs/1709.01515} {arXiv:1709.01515 [astro-ph.HE]}
  \BibitemShut {NoStop}%
\bibitem [{\citenamefont {{M{\"u}ller}}(2019)}]{2019ARNPS..69..253M}%
  \BibitemOpen
  \bibfield  {author} {\bibinfo {author} {\bibfnamefont {B.}~\bibnamefont
  {{M{\"u}ller}}},\ }\bibfield  {title} {\enquote {\bibinfo {title} {{Neutrino
  Emission as Diagnostics of Core-Collapse Supernovae}},}\ }\href {\doibase
  10.1146/annurev-nucl-101918-023434} {\bibfield  {journal} {\bibinfo
  {journal} {Annual Review of Nuclear and Particle Science}\ }\textbf {\bibinfo
  {volume} {69}},\ \bibinfo {pages} {253--278} (\bibinfo {year} {2019})},\
  \Eprint {http://arxiv.org/abs/1904.11067} {arXiv:1904.11067 [astro-ph.HE]}
  \BibitemShut {NoStop}%
\bibitem [{\citenamefont {{Kuroda}}\ \emph {et~al.}(2016)\citenamefont
  {{Kuroda}}, \citenamefont {{Kotake}},\ and\ \citenamefont
  {{Takiwaki}}}]{2016ApJ...829L..14K}%
  \BibitemOpen
  \bibfield  {author} {\bibinfo {author} {\bibfnamefont {Takami}\ \bibnamefont
  {{Kuroda}}}, \bibinfo {author} {\bibfnamefont {Kei}\ \bibnamefont
  {{Kotake}}}, \ and\ \bibinfo {author} {\bibfnamefont {Tomoya}\ \bibnamefont
  {{Takiwaki}}},\ }\bibfield  {title} {\enquote {\bibinfo {title} {{A New
  Gravitational-wave Signature from Standing Accretion Shock Instability in
  Supernovae}},}\ }\href {\doibase 10.3847/2041-8205/829/1/L14} {\bibfield
  {journal} {\bibinfo  {journal} {\apj}\ }\textbf {\bibinfo {volume} {829}},\
  \bibinfo {eid} {L14} (\bibinfo {year} {2016})}\BibitemShut {NoStop}%
\bibitem [{\citenamefont {{Andresen}}\ \emph {et~al.}(2017)\citenamefont
  {{Andresen}}, \citenamefont {{M{\"u}ller}}, \citenamefont {{M{\"u}ller}},\
  and\ \citenamefont {{Janka}}}]{2017MNRAS.468.2032A}%
  \BibitemOpen
  \bibfield  {author} {\bibinfo {author} {\bibfnamefont {H.}~\bibnamefont
  {{Andresen}}}, \bibinfo {author} {\bibfnamefont {B.}~\bibnamefont
  {{M{\"u}ller}}}, \bibinfo {author} {\bibfnamefont {E.}~\bibnamefont
  {{M{\"u}ller}}}, \ and\ \bibinfo {author} {\bibfnamefont {H.~Th.}\
  \bibnamefont {{Janka}}},\ }\bibfield  {title} {\enquote {\bibinfo {title}
  {{Gravitational wave signals from 3D neutrino hydrodynamics simulations of
  core-collapse supernovae}},}\ }\href {\doibase 10.1093/mnras/stx618}
  {\bibfield  {journal} {\bibinfo  {journal} {\mnras}\ }\textbf {\bibinfo
  {volume} {468}},\ \bibinfo {pages} {2032--2051} (\bibinfo {year}
  {2017})}\BibitemShut {NoStop}%
\bibitem [{\citenamefont {{Radice}}\ \emph {et~al.}(2019)\citenamefont
  {{Radice}}, \citenamefont {{Morozova}}, \citenamefont {{Burrows}},
  \citenamefont {{Vartanyan}},\ and\ \citenamefont
  {{Nagakura}}}]{2019ApJ...876L...9R}%
  \BibitemOpen
  \bibfield  {author} {\bibinfo {author} {\bibfnamefont {David}\ \bibnamefont
  {{Radice}}}, \bibinfo {author} {\bibfnamefont {Viktoriya}\ \bibnamefont
  {{Morozova}}}, \bibinfo {author} {\bibfnamefont {Adam}\ \bibnamefont
  {{Burrows}}}, \bibinfo {author} {\bibfnamefont {David}\ \bibnamefont
  {{Vartanyan}}}, \ and\ \bibinfo {author} {\bibfnamefont {Hiroki}\
  \bibnamefont {{Nagakura}}},\ }\bibfield  {title} {\enquote {\bibinfo {title}
  {{Characterizing the Gravitational Wave Signal from Core-collapse
  Supernovae}},}\ }\href {\doibase 10.3847/2041-8213/ab191a} {\bibfield
  {journal} {\bibinfo  {journal} {\apjl}\ }\textbf {\bibinfo {volume} {876}},\
  \bibinfo {eid} {L9} (\bibinfo {year} {2019})},\ \Eprint
  {http://arxiv.org/abs/1812.07703} {arXiv:1812.07703 [astro-ph.HE]}
  \BibitemShut {NoStop}%
\bibitem [{\citenamefont {{Mezzacappa}}\ \emph {et~al.}(2020)\citenamefont
  {{Mezzacappa}}, \citenamefont {{Marronetti}}, \citenamefont {{Landfield}},
  \citenamefont {{Lentz}}, \citenamefont {{Yakunin}}, \citenamefont {{Bruenn}},
  \citenamefont {{Hix}}, \citenamefont {{Messer}}, \citenamefont {{Endeve}},
  \citenamefont {{Blondin}},\ and\ \citenamefont
  {{Harris}}}]{2020PhRvD.102b3027M}%
  \BibitemOpen
  \bibfield  {author} {\bibinfo {author} {\bibfnamefont {Anthony}\ \bibnamefont
  {{Mezzacappa}}}, \bibinfo {author} {\bibfnamefont {Pedro}\ \bibnamefont
  {{Marronetti}}}, \bibinfo {author} {\bibfnamefont {Ryan~E.}\ \bibnamefont
  {{Landfield}}}, \bibinfo {author} {\bibfnamefont {Eric~J.}\ \bibnamefont
  {{Lentz}}}, \bibinfo {author} {\bibfnamefont {Konstantin~N.}\ \bibnamefont
  {{Yakunin}}}, \bibinfo {author} {\bibfnamefont {Stephen~W.}\ \bibnamefont
  {{Bruenn}}}, \bibinfo {author} {\bibfnamefont {W.~Raphael}\ \bibnamefont
  {{Hix}}}, \bibinfo {author} {\bibfnamefont {O.~E.~Bronson}\ \bibnamefont
  {{Messer}}}, \bibinfo {author} {\bibfnamefont {Eirik}\ \bibnamefont
  {{Endeve}}}, \bibinfo {author} {\bibfnamefont {John~M.}\ \bibnamefont
  {{Blondin}}}, \ and\ \bibinfo {author} {\bibfnamefont {J.~Austin}\
  \bibnamefont {{Harris}}},\ }\bibfield  {title} {\enquote {\bibinfo {title}
  {{Gravitational-wave signal of a core-collapse supernova explosion of a 15 M
  star}},}\ }\href {\doibase 10.1103/PhysRevD.102.023027} {\bibfield  {journal}
  {\bibinfo  {journal} {\prd}\ }\textbf {\bibinfo {volume} {102}},\ \bibinfo
  {eid} {023027} (\bibinfo {year} {2020})},\ \Eprint
  {http://arxiv.org/abs/2007.15099} {arXiv:2007.15099 [astro-ph.HE]}
  \BibitemShut {NoStop}%
\bibitem [{\citenamefont {{Jakobus}}\ \emph {et~al.}(2023)\citenamefont
  {{Jakobus}}, \citenamefont {{M{\"u}ller}}, \citenamefont {{Heger}},
  \citenamefont {{Zha}}, \citenamefont {{Powell}}, \citenamefont
  {{Motornenko}}, \citenamefont {{Steinheimer}},\ and\ \citenamefont
  {{Stoecker}}}]{2023arXiv230106515J}%
  \BibitemOpen
  \bibfield  {author} {\bibinfo {author} {\bibfnamefont {Pia}\ \bibnamefont
  {{Jakobus}}}, \bibinfo {author} {\bibfnamefont {Bernhard}\ \bibnamefont
  {{M{\"u}ller}}}, \bibinfo {author} {\bibfnamefont {Alexander}\ \bibnamefont
  {{Heger}}}, \bibinfo {author} {\bibfnamefont {Shuai}\ \bibnamefont {{Zha}}},
  \bibinfo {author} {\bibfnamefont {Jade}\ \bibnamefont {{Powell}}}, \bibinfo
  {author} {\bibfnamefont {Anton}\ \bibnamefont {{Motornenko}}}, \bibinfo
  {author} {\bibfnamefont {Jan}\ \bibnamefont {{Steinheimer}}}, \ and\ \bibinfo
  {author} {\bibfnamefont {Horst}\ \bibnamefont {{Stoecker}}},\ }\bibfield
  {title} {\enquote {\bibinfo {title} {{Gravitational Waves from a Core g-Mode
  in Supernovae as Probes of the High-Density Equation of State}},}\ }\href
  {\doibase 10.48550/arXiv.2301.06515} {\bibfield  {journal} {\bibinfo
  {journal} {arXiv e-prints}\ ,\ \bibinfo {eid} {arXiv:2301.06515}} (\bibinfo
  {year} {2023})},\ \Eprint {http://arxiv.org/abs/2301.06515} {arXiv:2301.06515
  [astro-ph.HE]} \BibitemShut {NoStop}%
\bibitem [{\citenamefont {{Fuller}}\ \emph {et~al.}(2015)\citenamefont
  {{Fuller}}, \citenamefont {{Klion}}, \citenamefont {{Abdikamalov}},\ and\
  \citenamefont {{Ott}}}]{2015MNRAS.450..414F}%
  \BibitemOpen
  \bibfield  {author} {\bibinfo {author} {\bibfnamefont {Jim}\ \bibnamefont
  {{Fuller}}}, \bibinfo {author} {\bibfnamefont {Hannah}\ \bibnamefont
  {{Klion}}}, \bibinfo {author} {\bibfnamefont {Ernazar}\ \bibnamefont
  {{Abdikamalov}}}, \ and\ \bibinfo {author} {\bibfnamefont {Christian~D.}\
  \bibnamefont {{Ott}}},\ }\bibfield  {title} {\enquote {\bibinfo {title}
  {{Supernova seismology: gravitational wave signatures of rapidly rotating
  core collapse}},}\ }\href {\doibase 10.1093/mnras/stv698} {\bibfield
  {journal} {\bibinfo  {journal} {\mnras}\ }\textbf {\bibinfo {volume} {450}},\
  \bibinfo {pages} {414--427} (\bibinfo {year} {2015})},\ \Eprint
  {http://arxiv.org/abs/1501.06951} {arXiv:1501.06951 [astro-ph.HE]}
  \BibitemShut {NoStop}%
\bibitem [{\citenamefont {{Sotani}}\ \emph {et~al.}(2017)\citenamefont
  {{Sotani}}, \citenamefont {{Kuroda}}, \citenamefont {{Takiwaki}},\ and\
  \citenamefont {{Kotake}}}]{2017PhRvD..96f3005S}%
  \BibitemOpen
  \bibfield  {author} {\bibinfo {author} {\bibfnamefont {Hajime}\ \bibnamefont
  {{Sotani}}}, \bibinfo {author} {\bibfnamefont {Takami}\ \bibnamefont
  {{Kuroda}}}, \bibinfo {author} {\bibfnamefont {Tomoya}\ \bibnamefont
  {{Takiwaki}}}, \ and\ \bibinfo {author} {\bibfnamefont {Kei}\ \bibnamefont
  {{Kotake}}},\ }\bibfield  {title} {\enquote {\bibinfo {title} {{Probing
  mass-radius relation of protoneutron stars from gravitational-wave
  asteroseismology}},}\ }\href {\doibase 10.1103/PhysRevD.96.063005} {\bibfield
   {journal} {\bibinfo  {journal} {\prd}\ }\textbf {\bibinfo {volume} {96}},\
  \bibinfo {eid} {063005} (\bibinfo {year} {2017})},\ \Eprint
  {http://arxiv.org/abs/1708.03738} {arXiv:1708.03738 [astro-ph.HE]}
  \BibitemShut {NoStop}%
\bibitem [{\citenamefont {{Torres-Forn{\'e}}}\ \emph
  {et~al.}(2018)\citenamefont {{Torres-Forn{\'e}}}, \citenamefont
  {{Cerd{\'a}-Dur{\'a}n}}, \citenamefont {{Passamonti}},\ and\ \citenamefont
  {{Font}}}]{2018MNRAS.474.5272T}%
  \BibitemOpen
  \bibfield  {author} {\bibinfo {author} {\bibfnamefont {Alejandro}\
  \bibnamefont {{Torres-Forn{\'e}}}}, \bibinfo {author} {\bibfnamefont {Pablo}\
  \bibnamefont {{Cerd{\'a}-Dur{\'a}n}}}, \bibinfo {author} {\bibfnamefont
  {Andrea}\ \bibnamefont {{Passamonti}}}, \ and\ \bibinfo {author}
  {\bibfnamefont {Jos{\'e}~A.}\ \bibnamefont {{Font}}},\ }\bibfield  {title}
  {\enquote {\bibinfo {title} {{Towards asteroseismology of core-collapse
  supernovae with gravitational-wave observations - I. Cowling
  approximation}},}\ }\href {\doibase 10.1093/mnras/stx3067} {\bibfield
  {journal} {\bibinfo  {journal} {\mnras}\ }\textbf {\bibinfo {volume} {474}},\
  \bibinfo {pages} {5272--5286} (\bibinfo {year} {2018})},\ \Eprint
  {http://arxiv.org/abs/1708.01920} {arXiv:1708.01920 [astro-ph.SR]}
  \BibitemShut {NoStop}%
\bibitem [{\citenamefont {{Morozova}}\ \emph {et~al.}(2018)\citenamefont
  {{Morozova}}, \citenamefont {{Radice}}, \citenamefont {{Burrows}},\ and\
  \citenamefont {{Vartanyan}}}]{2018ApJ...861...10M}%
  \BibitemOpen
  \bibfield  {author} {\bibinfo {author} {\bibfnamefont {V.}~\bibnamefont
  {{Morozova}}}, \bibinfo {author} {\bibfnamefont {D.}~\bibnamefont
  {{Radice}}}, \bibinfo {author} {\bibfnamefont {A.}~\bibnamefont {{Burrows}}},
  \ and\ \bibinfo {author} {\bibfnamefont {D.}~\bibnamefont {{Vartanyan}}},\
  }\bibfield  {title} {\enquote {\bibinfo {title} {{The Gravitational Wave
  Signal from Core-collapse Supernovae}},}\ }\href {\doibase
  10.3847/1538-4357/aac5f1} {\bibfield  {journal} {\bibinfo  {journal} {\apj}\
  }\textbf {\bibinfo {volume} {861}},\ \bibinfo {eid} {10} (\bibinfo {year}
  {2018})},\ \Eprint {http://arxiv.org/abs/1801.01914} {arXiv:1801.01914
  [astro-ph.HE]} \BibitemShut {NoStop}%
\bibitem [{\citenamefont {{Torres-Forn{\'e}}}\ \emph
  {et~al.}(2019{\natexlab{a}})\citenamefont {{Torres-Forn{\'e}}}, \citenamefont
  {{Cerd{\'a}-Dur{\'a}n}}, \citenamefont {{Passamonti}}, \citenamefont
  {{Obergaulinger}},\ and\ \citenamefont {{Font}}}]{2019MNRAS.482.3967T}%
  \BibitemOpen
  \bibfield  {author} {\bibinfo {author} {\bibfnamefont {Alejandro}\
  \bibnamefont {{Torres-Forn{\'e}}}}, \bibinfo {author} {\bibfnamefont {Pablo}\
  \bibnamefont {{Cerd{\'a}-Dur{\'a}n}}}, \bibinfo {author} {\bibfnamefont
  {Andrea}\ \bibnamefont {{Passamonti}}}, \bibinfo {author} {\bibfnamefont
  {Martin}\ \bibnamefont {{Obergaulinger}}}, \ and\ \bibinfo {author}
  {\bibfnamefont {Jos{\'e}~A.}\ \bibnamefont {{Font}}},\ }\bibfield  {title}
  {\enquote {\bibinfo {title} {{Towards asteroseismology of core-collapse
  supernovae with gravitational wave observations - II. Inclusion of space-time
  perturbations}},}\ }\href {\doibase 10.1093/mnras/sty2854} {\bibfield
  {journal} {\bibinfo  {journal} {\mnras}\ }\textbf {\bibinfo {volume} {482}},\
  \bibinfo {pages} {3967--3988} (\bibinfo {year} {2019}{\natexlab{a}})},\
  \Eprint {http://arxiv.org/abs/1806.11366} {arXiv:1806.11366 [astro-ph.HE]}
  \BibitemShut {NoStop}%
\bibitem [{\citenamefont {{Torres-Forn{\'e}}}\ \emph
  {et~al.}(2019{\natexlab{b}})\citenamefont {{Torres-Forn{\'e}}}, \citenamefont
  {{Cerd{\'a}-Dur{\'a}n}}, \citenamefont {{Obergaulinger}}, \citenamefont
  {{M{\"u}ller}},\ and\ \citenamefont {{Font}}}]{2019PhRvL.123e1102T}%
  \BibitemOpen
  \bibfield  {author} {\bibinfo {author} {\bibfnamefont {Alejandro}\
  \bibnamefont {{Torres-Forn{\'e}}}}, \bibinfo {author} {\bibfnamefont {Pablo}\
  \bibnamefont {{Cerd{\'a}-Dur{\'a}n}}}, \bibinfo {author} {\bibfnamefont
  {Martin}\ \bibnamefont {{Obergaulinger}}}, \bibinfo {author} {\bibfnamefont
  {Bernhard}\ \bibnamefont {{M{\"u}ller}}}, \ and\ \bibinfo {author}
  {\bibfnamefont {Jos{\'e}~A.}\ \bibnamefont {{Font}}},\ }\bibfield  {title}
  {\enquote {\bibinfo {title} {{Universal Relations for Gravitational-Wave
  Asteroseismology of Protoneutron Stars}},}\ }\href {\doibase
  10.1103/PhysRevLett.123.051102} {\bibfield  {journal} {\bibinfo  {journal}
  {\prl}\ }\textbf {\bibinfo {volume} {123}},\ \bibinfo {eid} {051102}
  (\bibinfo {year} {2019}{\natexlab{b}})},\ \Eprint
  {http://arxiv.org/abs/1902.10048} {arXiv:1902.10048 [gr-qc]} \BibitemShut
  {NoStop}%
\bibitem [{\citenamefont {{Bizouard}}\ \emph {et~al.}(2021)\citenamefont
  {{Bizouard}}, \citenamefont {{Maturana-Russel}}, \citenamefont
  {{Torres-Forn{\'e}}}, \citenamefont {{Obergaulinger}}, \citenamefont
  {{Cerd{\'a}-Dur{\'a}n}}, \citenamefont {{Christensen}}, \citenamefont
  {{Font}},\ and\ \citenamefont {{Meyer}}}]{2021PhRvD.103f3006B}%
  \BibitemOpen
  \bibfield  {author} {\bibinfo {author} {\bibfnamefont {Marie-Anne}\
  \bibnamefont {{Bizouard}}}, \bibinfo {author} {\bibfnamefont {Patricio}\
  \bibnamefont {{Maturana-Russel}}}, \bibinfo {author} {\bibfnamefont
  {Alejandro}\ \bibnamefont {{Torres-Forn{\'e}}}}, \bibinfo {author}
  {\bibfnamefont {Martin}\ \bibnamefont {{Obergaulinger}}}, \bibinfo {author}
  {\bibfnamefont {Pablo}\ \bibnamefont {{Cerd{\'a}-Dur{\'a}n}}}, \bibinfo
  {author} {\bibfnamefont {Nelson}\ \bibnamefont {{Christensen}}}, \bibinfo
  {author} {\bibfnamefont {Jos{\'e}~A.}\ \bibnamefont {{Font}}}, \ and\
  \bibinfo {author} {\bibfnamefont {Renate}\ \bibnamefont {{Meyer}}},\
  }\bibfield  {title} {\enquote {\bibinfo {title} {{Inference of protoneutron
  star properties from gravitational-wave data in core-collapse supernovae}},}\
  }\href {\doibase 10.1103/PhysRevD.103.063006} {\bibfield  {journal} {\bibinfo
   {journal} {\prd}\ }\textbf {\bibinfo {volume} {103}},\ \bibinfo {eid}
  {063006} (\bibinfo {year} {2021})},\ \Eprint
  {http://arxiv.org/abs/2012.00846} {arXiv:2012.00846 [gr-qc]} \BibitemShut
  {NoStop}%
\bibitem [{\citenamefont {{Bruel}}\ \emph {et~al.}(2023)\citenamefont
  {{Bruel}}, \citenamefont {{Bizouard}}, \citenamefont {{Obergaulinger}},
  \citenamefont {{Maturana-Russel}}, \citenamefont {{Torres-Forn{\'e}}},
  \citenamefont {{Cerd{\'a}-Dur{\'a}n}}, \citenamefont {{Christensen}},
  \citenamefont {{Font}},\ and\ \citenamefont {{Meyer}}}]{2023PhRvD.107h3029B}%
  \BibitemOpen
  \bibfield  {author} {\bibinfo {author} {\bibfnamefont {Tristan}\ \bibnamefont
  {{Bruel}}}, \bibinfo {author} {\bibfnamefont {Marie-Anne}\ \bibnamefont
  {{Bizouard}}}, \bibinfo {author} {\bibfnamefont {Martin}\ \bibnamefont
  {{Obergaulinger}}}, \bibinfo {author} {\bibfnamefont {Patricio}\ \bibnamefont
  {{Maturana-Russel}}}, \bibinfo {author} {\bibfnamefont {Alejandro}\
  \bibnamefont {{Torres-Forn{\'e}}}}, \bibinfo {author} {\bibfnamefont {Pablo}\
  \bibnamefont {{Cerd{\'a}-Dur{\'a}n}}}, \bibinfo {author} {\bibfnamefont
  {Nelson}\ \bibnamefont {{Christensen}}}, \bibinfo {author} {\bibfnamefont
  {Jos{\'e}~A.}\ \bibnamefont {{Font}}}, \ and\ \bibinfo {author}
  {\bibfnamefont {Renate}\ \bibnamefont {{Meyer}}},\ }\bibfield  {title}
  {\enquote {\bibinfo {title} {{Inference of protoneutron star properties in
  core-collapse supernovae from a gravitational-wave detector network}},}\
  }\href {\doibase 10.1103/PhysRevD.107.083029} {\bibfield  {journal} {\bibinfo
   {journal} {\prd}\ }\textbf {\bibinfo {volume} {107}},\ \bibinfo {eid}
  {083029} (\bibinfo {year} {2023})},\ \Eprint
  {http://arxiv.org/abs/2301.10019} {arXiv:2301.10019 [astro-ph.HE]}
  \BibitemShut {NoStop}%
\bibitem [{\citenamefont {{Nakamura}}\ \emph {et~al.}(2016)\citenamefont
  {{Nakamura}}, \citenamefont {{Horiuchi}}, \citenamefont {{Tanaka}},
  \citenamefont {{Hayama}}, \citenamefont {{Takiwaki}},\ and\ \citenamefont
  {{Kotake}}}]{2016MNRAS.461.3296N}%
  \BibitemOpen
  \bibfield  {author} {\bibinfo {author} {\bibfnamefont {Ko}~\bibnamefont
  {{Nakamura}}}, \bibinfo {author} {\bibfnamefont {Shunsaku}\ \bibnamefont
  {{Horiuchi}}}, \bibinfo {author} {\bibfnamefont {Masaomi}\ \bibnamefont
  {{Tanaka}}}, \bibinfo {author} {\bibfnamefont {Kazuhiro}\ \bibnamefont
  {{Hayama}}}, \bibinfo {author} {\bibfnamefont {Tomoya}\ \bibnamefont
  {{Takiwaki}}}, \ and\ \bibinfo {author} {\bibfnamefont {Kei}\ \bibnamefont
  {{Kotake}}},\ }\bibfield  {title} {\enquote {\bibinfo {title}
  {{Multimessenger signals of long-term core-collapse supernova simulations:
  synergetic observation strategies}},}\ }\href {\doibase
  10.1093/mnras/stw1453} {\bibfield  {journal} {\bibinfo  {journal} {\mnras}\
  }\textbf {\bibinfo {volume} {461}},\ \bibinfo {pages} {3296--3313} (\bibinfo
  {year} {2016})},\ \Eprint {http://arxiv.org/abs/1602.03028} {arXiv:1602.03028
  [astro-ph.HE]} \BibitemShut {NoStop}%
\bibitem [{\citenamefont {{Kuroda}}\ \emph {et~al.}(2017)\citenamefont
  {{Kuroda}}, \citenamefont {{Kotake}}, \citenamefont {{Hayama}},\ and\
  \citenamefont {{Takiwaki}}}]{2017ApJ...851...62K}%
  \BibitemOpen
  \bibfield  {author} {\bibinfo {author} {\bibfnamefont {Takami}\ \bibnamefont
  {{Kuroda}}}, \bibinfo {author} {\bibfnamefont {Kei}\ \bibnamefont
  {{Kotake}}}, \bibinfo {author} {\bibfnamefont {Kazuhiro}\ \bibnamefont
  {{Hayama}}}, \ and\ \bibinfo {author} {\bibfnamefont {Tomoya}\ \bibnamefont
  {{Takiwaki}}},\ }\bibfield  {title} {\enquote {\bibinfo {title} {{Correlated
  Signatures of Gravitational-wave and Neutrino Emission in Three-dimensional
  General-relativistic Core-collapse Supernova Simulations}},}\ }\href
  {\doibase 10.3847/1538-4357/aa988d} {\bibfield  {journal} {\bibinfo
  {journal} {\apj}\ }\textbf {\bibinfo {volume} {851}},\ \bibinfo {eid} {62}
  (\bibinfo {year} {2017})},\ \Eprint {http://arxiv.org/abs/1708.05252}
  {arXiv:1708.05252 [astro-ph.HE]} \BibitemShut {NoStop}%
\bibitem [{\citenamefont {{Warren}}\ \emph {et~al.}(2020)\citenamefont
  {{Warren}}, \citenamefont {{Couch}}, \citenamefont {{O'Connor}},\ and\
  \citenamefont {{Morozova}}}]{2020ApJ...898..139W}%
  \BibitemOpen
  \bibfield  {author} {\bibinfo {author} {\bibfnamefont {MacKenzie~L.}\
  \bibnamefont {{Warren}}}, \bibinfo {author} {\bibfnamefont {Sean~M.}\
  \bibnamefont {{Couch}}}, \bibinfo {author} {\bibfnamefont {Evan~P.}\
  \bibnamefont {{O'Connor}}}, \ and\ \bibinfo {author} {\bibfnamefont
  {Viktoriya}\ \bibnamefont {{Morozova}}},\ }\bibfield  {title} {\enquote
  {\bibinfo {title} {{Constraining Properties of the Next Nearby Core-collapse
  Supernova with Multimessenger Signals}},}\ }\href {\doibase
  10.3847/1538-4357/ab97b7} {\bibfield  {journal} {\bibinfo  {journal} {\apj}\
  }\textbf {\bibinfo {volume} {898}},\ \bibinfo {eid} {139} (\bibinfo {year}
  {2020})},\ \Eprint {http://arxiv.org/abs/1912.03328} {arXiv:1912.03328
  [astro-ph.HE]} \BibitemShut {NoStop}%
\bibitem [{\citenamefont {{Szczepa{\'n}czyk}}\ \emph
  {et~al.}(2021)\citenamefont {{Szczepa{\'n}czyk}}, \citenamefont {{Antelis}},
  \citenamefont {{Benjamin}}, \citenamefont {{Cavagli{\`a}}}, \citenamefont
  {{Gondek-Rosi{\'n}ska}}, \citenamefont {{Hansen}}, \citenamefont
  {{Klimenko}}, \citenamefont {{Morales}}, \citenamefont {{Moreno}},
  \citenamefont {{Mukherjee}}, \citenamefont {{Nurbek}}, \citenamefont
  {{Powell}}, \citenamefont {{Singh}}, \citenamefont {{Sitmukhambetov}},
  \citenamefont {{Szewczyk}}, \citenamefont {{Valdez}}, \citenamefont
  {{Vedovato}}, \citenamefont {{Westhouse}}, \citenamefont {{Zanolin}},\ and\
  \citenamefont {{Zheng}}}]{2021PhRvD.104j2002S}%
  \BibitemOpen
  \bibfield  {author} {\bibinfo {author} {\bibfnamefont {Marek~J.}\
  \bibnamefont {{Szczepa{\'n}czyk}}}, \bibinfo {author} {\bibfnamefont
  {Javier~M.}\ \bibnamefont {{Antelis}}}, \bibinfo {author} {\bibfnamefont
  {Michael}\ \bibnamefont {{Benjamin}}}, \bibinfo {author} {\bibfnamefont
  {Marco}\ \bibnamefont {{Cavagli{\`a}}}}, \bibinfo {author} {\bibfnamefont
  {Dorota}\ \bibnamefont {{Gondek-Rosi{\'n}ska}}}, \bibinfo {author}
  {\bibfnamefont {Travis}\ \bibnamefont {{Hansen}}}, \bibinfo {author}
  {\bibfnamefont {Sergey}\ \bibnamefont {{Klimenko}}}, \bibinfo {author}
  {\bibfnamefont {Manuel~D.}\ \bibnamefont {{Morales}}}, \bibinfo {author}
  {\bibfnamefont {Claudia}\ \bibnamefont {{Moreno}}}, \bibinfo {author}
  {\bibfnamefont {Soma}\ \bibnamefont {{Mukherjee}}}, \bibinfo {author}
  {\bibfnamefont {Gaukhar}\ \bibnamefont {{Nurbek}}}, \bibinfo {author}
  {\bibfnamefont {Jade}\ \bibnamefont {{Powell}}}, \bibinfo {author}
  {\bibfnamefont {Neha}\ \bibnamefont {{Singh}}}, \bibinfo {author}
  {\bibfnamefont {Satzhan}\ \bibnamefont {{Sitmukhambetov}}}, \bibinfo {author}
  {\bibfnamefont {Pawe{\l}}\ \bibnamefont {{Szewczyk}}}, \bibinfo {author}
  {\bibfnamefont {Oscar}\ \bibnamefont {{Valdez}}}, \bibinfo {author}
  {\bibfnamefont {Gabriele}\ \bibnamefont {{Vedovato}}}, \bibinfo {author}
  {\bibfnamefont {Jonathan}\ \bibnamefont {{Westhouse}}}, \bibinfo {author}
  {\bibfnamefont {Michele}\ \bibnamefont {{Zanolin}}}, \ and\ \bibinfo {author}
  {\bibfnamefont {Yanyan}\ \bibnamefont {{Zheng}}},\ }\bibfield  {title}
  {\enquote {\bibinfo {title} {{Detecting and reconstructing gravitational
  waves from the next galactic core-collapse supernova in the advanced detector
  era}},}\ }\href {\doibase 10.1103/PhysRevD.104.102002} {\bibfield  {journal}
  {\bibinfo  {journal} {\prd}\ }\textbf {\bibinfo {volume} {104}},\ \bibinfo
  {eid} {102002} (\bibinfo {year} {2021})},\ \Eprint
  {http://arxiv.org/abs/2104.06462} {arXiv:2104.06462 [astro-ph.HE]}
  \BibitemShut {NoStop}%
\bibitem [{\citenamefont {{Drago}}\ \emph {et~al.}(2023)\citenamefont
  {{Drago}}, \citenamefont {{Andresen}}, \citenamefont {{Di Palma}},
  \citenamefont {{Tamborra}},\ and\ \citenamefont
  {{Torres-Forn{\'e}}}}]{2023arXiv230507688D}%
  \BibitemOpen
  \bibfield  {author} {\bibinfo {author} {\bibfnamefont {Marco}\ \bibnamefont
  {{Drago}}}, \bibinfo {author} {\bibfnamefont {Haakon}\ \bibnamefont
  {{Andresen}}}, \bibinfo {author} {\bibfnamefont {Irene}\ \bibnamefont {{Di
  Palma}}}, \bibinfo {author} {\bibfnamefont {Irene}\ \bibnamefont
  {{Tamborra}}}, \ and\ \bibinfo {author} {\bibfnamefont {Alejandro}\
  \bibnamefont {{Torres-Forn{\'e}}}},\ }\bibfield  {title} {\enquote {\bibinfo
  {title} {{Multi-messenger observations of core-collapse supernovae:
  Exploiting the standing accretion shock instability}},}\ }\href {\doibase
  10.48550/arXiv.2305.07688} {\bibfield  {journal} {\bibinfo  {journal} {arXiv
  e-prints}\ ,\ \bibinfo {eid} {arXiv:2305.07688}} (\bibinfo {year} {2023})},\
  \Eprint {http://arxiv.org/abs/2305.07688} {arXiv:2305.07688 [astro-ph.HE]}
  \BibitemShut {NoStop}%
\bibitem [{\citenamefont {{Abbott}}\ and\ \citenamefont
  {et~al.}(2020)}]{2020PhRvD.101h4002A}%
  \BibitemOpen
  \bibfield  {author} {\bibinfo {author} {\bibfnamefont {B.~P.}\ \bibnamefont
  {{Abbott}}}\ and\ \bibinfo {author} {\bibnamefont {et~al.}},\ }\bibfield
  {title} {\enquote {\bibinfo {title} {{Optically targeted search for
  gravitational waves emitted by core-collapse supernovae during the first and
  second observing runs of advanced LIGO and advanced Virgo}},}\ }\href
  {\doibase 10.1103/PhysRevD.101.084002} {\bibfield  {journal} {\bibinfo
  {journal} {\prd}\ }\textbf {\bibinfo {volume} {101}},\ \bibinfo {eid}
  {084002} (\bibinfo {year} {2020})},\ \Eprint
  {http://arxiv.org/abs/1908.03584} {arXiv:1908.03584 [astro-ph.HE]}
  \BibitemShut {NoStop}%
\bibitem [{\citenamefont {{Powell}}\ and\ \citenamefont
  {{M{\"u}ller}}(2020)}]{2020MNRAS.494.4665P}%
  \BibitemOpen
  \bibfield  {author} {\bibinfo {author} {\bibfnamefont {Jade}\ \bibnamefont
  {{Powell}}}\ and\ \bibinfo {author} {\bibfnamefont {Bernhard}\ \bibnamefont
  {{M{\"u}ller}}},\ }\bibfield  {title} {\enquote {\bibinfo {title}
  {{Three-dimensional core-collapse supernova simulations of massive and
  rotating progenitors}},}\ }\href {\doibase 10.1093/mnras/staa1048} {\bibfield
   {journal} {\bibinfo  {journal} {\mnras}\ }\textbf {\bibinfo {volume}
  {494}},\ \bibinfo {pages} {4665--4675} (\bibinfo {year} {2020})},\ \Eprint
  {http://arxiv.org/abs/2002.10115} {arXiv:2002.10115 [astro-ph.HE]}
  \BibitemShut {NoStop}%
\bibitem [{\citenamefont {{Mezzacappa}}\ \emph {et~al.}(2023)\citenamefont
  {{Mezzacappa}}, \citenamefont {{Marronetti}}, \citenamefont {{Landfield}},
  \citenamefont {{Lentz}}, \citenamefont {{Murphy}}, \citenamefont {{Hix}},
  \citenamefont {{Harris}}, \citenamefont {{Bruenn}}, \citenamefont
  {{Blondin}}, \citenamefont {{Bronson Messer}}, \citenamefont {{Casanova}},\
  and\ \citenamefont {{Kronzer}}}]{2023PhRvD.107d3008M}%
  \BibitemOpen
  \bibfield  {author} {\bibinfo {author} {\bibfnamefont {Anthony}\ \bibnamefont
  {{Mezzacappa}}}, \bibinfo {author} {\bibfnamefont {Pedro}\ \bibnamefont
  {{Marronetti}}}, \bibinfo {author} {\bibfnamefont {Ryan~E.}\ \bibnamefont
  {{Landfield}}}, \bibinfo {author} {\bibfnamefont {Eric~J.}\ \bibnamefont
  {{Lentz}}}, \bibinfo {author} {\bibfnamefont {R.~Daniel}\ \bibnamefont
  {{Murphy}}}, \bibinfo {author} {\bibfnamefont {W.~Raphael}\ \bibnamefont
  {{Hix}}}, \bibinfo {author} {\bibfnamefont {J.~Austin}\ \bibnamefont
  {{Harris}}}, \bibinfo {author} {\bibfnamefont {Stephen~W.}\ \bibnamefont
  {{Bruenn}}}, \bibinfo {author} {\bibfnamefont {John~M.}\ \bibnamefont
  {{Blondin}}}, \bibinfo {author} {\bibfnamefont {O.~E.}\ \bibnamefont
  {{Bronson Messer}}}, \bibinfo {author} {\bibfnamefont {Jordi}\ \bibnamefont
  {{Casanova}}}, \ and\ \bibinfo {author} {\bibfnamefont {Luke~L.}\
  \bibnamefont {{Kronzer}}},\ }\bibfield  {title} {\enquote {\bibinfo {title}
  {{Core collapse supernova gravitational wave emission for progenitors of 9.6,
  15, and 25 M$_{{\ensuremath{\odot}}}$}},}\ }\href {\doibase
  10.1103/PhysRevD.107.043008} {\bibfield  {journal} {\bibinfo  {journal}
  {\prd}\ }\textbf {\bibinfo {volume} {107}},\ \bibinfo {eid} {043008}
  (\bibinfo {year} {2023})},\ \Eprint {http://arxiv.org/abs/2208.10643}
  {arXiv:2208.10643 [astro-ph.SR]} \BibitemShut {NoStop}%
\bibitem [{\citenamefont {{Dighe}}\ and\ \citenamefont
  {{Smirnov}}(2000)}]{2000PhRvD..62c3007D}%
  \BibitemOpen
  \bibfield  {author} {\bibinfo {author} {\bibfnamefont {Amol~S.}\ \bibnamefont
  {{Dighe}}}\ and\ \bibinfo {author} {\bibfnamefont {Alexei~Yu.}\ \bibnamefont
  {{Smirnov}}},\ }\bibfield  {title} {\enquote {\bibinfo {title} {{Identifying
  the neutrino mass spectrum from a supernova neutrino burst}},}\ }\href
  {\doibase 10.1103/PhysRevD.62.033007} {\bibfield  {journal} {\bibinfo
  {journal} {\prd}\ }\textbf {\bibinfo {volume} {62}},\ \bibinfo {eid} {033007}
  (\bibinfo {year} {2000})},\ \Eprint {http://arxiv.org/abs/hep-ph/9907423}
  {arXiv:hep-ph/9907423 [hep-ph]} \BibitemShut {NoStop}%
\bibitem [{\citenamefont {{Chakraborty}}\ \emph {et~al.}(2016)\citenamefont
  {{Chakraborty}}, \citenamefont {{Hansen}}, \citenamefont {{Izaguirre}},\ and\
  \citenamefont {{Raffelt}}}]{2016NuPhB.908..366C}%
  \BibitemOpen
  \bibfield  {author} {\bibinfo {author} {\bibfnamefont {Sovan}\ \bibnamefont
  {{Chakraborty}}}, \bibinfo {author} {\bibfnamefont {Rasmus}\ \bibnamefont
  {{Hansen}}}, \bibinfo {author} {\bibfnamefont {Ignacio}\ \bibnamefont
  {{Izaguirre}}}, \ and\ \bibinfo {author} {\bibfnamefont {Georg}\ \bibnamefont
  {{Raffelt}}},\ }\bibfield  {title} {\enquote {\bibinfo {title} {{Collective
  neutrino flavor conversion: Recent developments}},}\ }\href {\doibase
  10.1016/j.nuclphysb.2016.02.012} {\bibfield  {journal} {\bibinfo  {journal}
  {Nuclear Physics B}\ }\textbf {\bibinfo {volume} {908}},\ \bibinfo {pages}
  {366--381} (\bibinfo {year} {2016})},\ \Eprint
  {http://arxiv.org/abs/1602.02766} {arXiv:1602.02766 [hep-ph]} \BibitemShut
  {NoStop}%
\bibitem [{\citenamefont {{Tamborra}}\ and\ \citenamefont
  {{Shalgar}}(2021)}]{2021ARNPS..71..165T}%
  \BibitemOpen
  \bibfield  {author} {\bibinfo {author} {\bibfnamefont {Irene}\ \bibnamefont
  {{Tamborra}}}\ and\ \bibinfo {author} {\bibfnamefont {Shashank}\ \bibnamefont
  {{Shalgar}}},\ }\bibfield  {title} {\enquote {\bibinfo {title} {{New
  Developments in Flavor Evolution of a Dense Neutrino Gas}},}\ }\href
  {\doibase 10.1146/annurev-nucl-102920-050505} {\bibfield  {journal} {\bibinfo
   {journal} {Annual Review of Nuclear and Particle Science}\ }\textbf
  {\bibinfo {volume} {71}},\ \bibinfo {pages} {165--188} (\bibinfo {year}
  {2021})},\ \Eprint {http://arxiv.org/abs/2011.01948} {arXiv:2011.01948
  [astro-ph.HE]} \BibitemShut {NoStop}%
\bibitem [{\citenamefont {{Capozzi}}\ and\ \citenamefont
  {{Saviano}}(2022)}]{2022Univ....8...94C}%
  \BibitemOpen
  \bibfield  {author} {\bibinfo {author} {\bibfnamefont {Francesco}\
  \bibnamefont {{Capozzi}}}\ and\ \bibinfo {author} {\bibfnamefont {Ninetta}\
  \bibnamefont {{Saviano}}},\ }\bibfield  {title} {\enquote {\bibinfo {title}
  {{Neutrino Flavor Conversions in High-Density Astrophysical and Cosmological
  Environments}},}\ }\href {\doibase 10.3390/universe8020094} {\bibfield
  {journal} {\bibinfo  {journal} {Universe}\ }\textbf {\bibinfo {volume} {8}},\
  \bibinfo {pages} {94} (\bibinfo {year} {2022})},\ \Eprint
  {http://arxiv.org/abs/2202.02494} {arXiv:2202.02494 [hep-ph]} \BibitemShut
  {NoStop}%
\bibitem [{\citenamefont {{Richers}}\ and\ \citenamefont
  {{Sen}}(2022)}]{2022arXiv220703561R}%
  \BibitemOpen
  \bibfield  {author} {\bibinfo {author} {\bibfnamefont {Sherwood}\
  \bibnamefont {{Richers}}}\ and\ \bibinfo {author} {\bibfnamefont {Manibrata}\
  \bibnamefont {{Sen}}},\ }\bibfield  {title} {\enquote {\bibinfo {title}
  {{Fast Flavor Transformations}},}\ }\href@noop {} {\bibfield  {journal}
  {\bibinfo  {journal} {arXiv e-prints}\ ,\ \bibinfo {eid} {arXiv:2207.03561}}
  (\bibinfo {year} {2022})},\ \Eprint {http://arxiv.org/abs/2207.03561}
  {arXiv:2207.03561 [astro-ph.HE]} \BibitemShut {NoStop}%
\bibitem [{\citenamefont {{Nagakura}}\ \emph {et~al.}(2019)\citenamefont
  {{Nagakura}}, \citenamefont {{Morinaga}}, \citenamefont {{Kato}},\ and\
  \citenamefont {{Yamada}}}]{2019ApJ...886..139N}%
  \BibitemOpen
  \bibfield  {author} {\bibinfo {author} {\bibfnamefont {Hiroki}\ \bibnamefont
  {{Nagakura}}}, \bibinfo {author} {\bibfnamefont {Taiki}\ \bibnamefont
  {{Morinaga}}}, \bibinfo {author} {\bibfnamefont {Chinami}\ \bibnamefont
  {{Kato}}}, \ and\ \bibinfo {author} {\bibfnamefont {Shoichi}\ \bibnamefont
  {{Yamada}}},\ }\bibfield  {title} {\enquote {\bibinfo {title} {{Fast-pairwise
  Collective Neutrino Oscillations Associated with Asymmetric Neutrino
  Emissions in Core-collapse Supernovae}},}\ }\href {\doibase
  10.3847/1538-4357/ab4cf2} {\bibfield  {journal} {\bibinfo  {journal} {\apj}\
  }\textbf {\bibinfo {volume} {886}},\ \bibinfo {eid} {139} (\bibinfo {year}
  {2019})},\ \Eprint {http://arxiv.org/abs/1910.04288} {arXiv:1910.04288
  [astro-ph.HE]} \BibitemShut {NoStop}%
\bibitem [{\citenamefont {{Shalgar}}\ and\ \citenamefont
  {{Tamborra}}(2019)}]{2019ApJ...883...80S}%
  \BibitemOpen
  \bibfield  {author} {\bibinfo {author} {\bibfnamefont {Shashank}\
  \bibnamefont {{Shalgar}}}\ and\ \bibinfo {author} {\bibfnamefont {Irene}\
  \bibnamefont {{Tamborra}}},\ }\bibfield  {title} {\enquote {\bibinfo {title}
  {{On the Occurrence of Crossings between the Angular Distributions of
  Electron Neutrinos and Antineutrinos in the Supernova Core}},}\ }\href
  {\doibase 10.3847/1538-4357/ab38ba} {\bibfield  {journal} {\bibinfo
  {journal} {\apj}\ }\textbf {\bibinfo {volume} {883}},\ \bibinfo {eid} {80}
  (\bibinfo {year} {2019})},\ \Eprint {http://arxiv.org/abs/1904.07236}
  {arXiv:1904.07236 [astro-ph.HE]} \BibitemShut {NoStop}%
\bibitem [{\citenamefont {{Delfan Azari}}\ \emph {et~al.}(2020)\citenamefont
  {{Delfan Azari}}, \citenamefont {{Yamada}}, \citenamefont {{Morinaga}},
  \citenamefont {{Nagakura}}, \citenamefont {{Furusawa}}, \citenamefont
  {{Harada}}, \citenamefont {{Okawa}}, \citenamefont {{Iwakami}},\ and\
  \citenamefont {{Sumiyoshi}}}]{2020PhRvD.101b3018D}%
  \BibitemOpen
  \bibfield  {author} {\bibinfo {author} {\bibfnamefont {Milad}\ \bibnamefont
  {{Delfan Azari}}}, \bibinfo {author} {\bibfnamefont {Shoichi}\ \bibnamefont
  {{Yamada}}}, \bibinfo {author} {\bibfnamefont {Taiki}\ \bibnamefont
  {{Morinaga}}}, \bibinfo {author} {\bibfnamefont {Hiroki}\ \bibnamefont
  {{Nagakura}}}, \bibinfo {author} {\bibfnamefont {Shun}\ \bibnamefont
  {{Furusawa}}}, \bibinfo {author} {\bibfnamefont {Akira}\ \bibnamefont
  {{Harada}}}, \bibinfo {author} {\bibfnamefont {Hirotada}\ \bibnamefont
  {{Okawa}}}, \bibinfo {author} {\bibfnamefont {Wakana}\ \bibnamefont
  {{Iwakami}}}, \ and\ \bibinfo {author} {\bibfnamefont {Kohsuke}\ \bibnamefont
  {{Sumiyoshi}}},\ }\bibfield  {title} {\enquote {\bibinfo {title} {{Fast
  collective neutrino oscillations inside the neutrino sphere in core-collapse
  supernovae}},}\ }\href {\doibase 10.1103/PhysRevD.101.023018} {\bibfield
  {journal} {\bibinfo  {journal} {\prd}\ }\textbf {\bibinfo {volume} {101}},\
  \bibinfo {eid} {023018} (\bibinfo {year} {2020})},\ \Eprint
  {http://arxiv.org/abs/1910.06176} {arXiv:1910.06176 [astro-ph.HE]}
  \BibitemShut {NoStop}%
\bibitem [{\citenamefont {{Morinaga}}\ \emph {et~al.}(2020)\citenamefont
  {{Morinaga}}, \citenamefont {{Nagakura}}, \citenamefont {{Kato}},\ and\
  \citenamefont {{Yamada}}}]{2020PhRvR...2a2046M}%
  \BibitemOpen
  \bibfield  {author} {\bibinfo {author} {\bibfnamefont {Taiki}\ \bibnamefont
  {{Morinaga}}}, \bibinfo {author} {\bibfnamefont {Hiroki}\ \bibnamefont
  {{Nagakura}}}, \bibinfo {author} {\bibfnamefont {Chinami}\ \bibnamefont
  {{Kato}}}, \ and\ \bibinfo {author} {\bibfnamefont {Shoichi}\ \bibnamefont
  {{Yamada}}},\ }\bibfield  {title} {\enquote {\bibinfo {title} {{Fast
  neutrino-flavor conversion in the preshock region of core-collapse
  supernovae}},}\ }\href {\doibase 10.1103/PhysRevResearch.2.012046} {\bibfield
   {journal} {\bibinfo  {journal} {Physical Review Research}\ }\textbf
  {\bibinfo {volume} {2}},\ \bibinfo {eid} {012046} (\bibinfo {year} {2020})},\
  \Eprint {http://arxiv.org/abs/1909.13131} {arXiv:1909.13131 [astro-ph.HE]}
  \BibitemShut {NoStop}%
\bibitem [{\citenamefont {{Nagakura}}\ \emph
  {et~al.}(2021{\natexlab{a}})\citenamefont {{Nagakura}}, \citenamefont
  {{Burrows}}, \citenamefont {{Johns}},\ and\ \citenamefont
  {{Fuller}}}]{2021PhRvD.104h3025N}%
  \BibitemOpen
  \bibfield  {author} {\bibinfo {author} {\bibfnamefont {Hiroki}\ \bibnamefont
  {{Nagakura}}}, \bibinfo {author} {\bibfnamefont {Adam}\ \bibnamefont
  {{Burrows}}}, \bibinfo {author} {\bibfnamefont {Lucas}\ \bibnamefont
  {{Johns}}}, \ and\ \bibinfo {author} {\bibfnamefont {George~M.}\ \bibnamefont
  {{Fuller}}},\ }\bibfield  {title} {\enquote {\bibinfo {title} {{Where, when,
  and why: Occurrence of fast-pairwise collective neutrino oscillation in
  three-dimensional core-collapse supernova models}},}\ }\href {\doibase
  10.1103/PhysRevD.104.083025} {\bibfield  {journal} {\bibinfo  {journal}
  {\prd}\ }\textbf {\bibinfo {volume} {104}},\ \bibinfo {eid} {083025}
  (\bibinfo {year} {2021}{\natexlab{a}})},\ \Eprint
  {http://arxiv.org/abs/2108.07281} {arXiv:2108.07281 [astro-ph.HE]}
  \BibitemShut {NoStop}%
\bibitem [{\citenamefont {{Abbar}}\ \emph {et~al.}(2021)\citenamefont
  {{Abbar}}, \citenamefont {{Capozzi}}, \citenamefont {{Glas}}, \citenamefont
  {{Janka}},\ and\ \citenamefont {{Tamborra}}}]{2021PhRvD.103f3033A}%
  \BibitemOpen
  \bibfield  {author} {\bibinfo {author} {\bibfnamefont {Sajad}\ \bibnamefont
  {{Abbar}}}, \bibinfo {author} {\bibfnamefont {Francesco}\ \bibnamefont
  {{Capozzi}}}, \bibinfo {author} {\bibfnamefont {Robert}\ \bibnamefont
  {{Glas}}}, \bibinfo {author} {\bibfnamefont {H.~Thomas}\ \bibnamefont
  {{Janka}}}, \ and\ \bibinfo {author} {\bibfnamefont {Irene}\ \bibnamefont
  {{Tamborra}}},\ }\bibfield  {title} {\enquote {\bibinfo {title} {{On the
  characteristics of fast neutrino flavor instabilities in three-dimensional
  core-collapse supernova models}},}\ }\href {\doibase
  10.1103/PhysRevD.103.063033} {\bibfield  {journal} {\bibinfo  {journal}
  {\prd}\ }\textbf {\bibinfo {volume} {103}},\ \bibinfo {eid} {063033}
  (\bibinfo {year} {2021})},\ \Eprint {http://arxiv.org/abs/2012.06594}
  {arXiv:2012.06594 [astro-ph.HE]} \BibitemShut {NoStop}%
\bibitem [{\citenamefont {{Capozzi}}\ \emph {et~al.}(2021)\citenamefont
  {{Capozzi}}, \citenamefont {{Abbar}}, \citenamefont {{Bollig}},\ and\
  \citenamefont {{Janka}}}]{2021PhRvD.103f3013C}%
  \BibitemOpen
  \bibfield  {author} {\bibinfo {author} {\bibfnamefont {Francesco}\
  \bibnamefont {{Capozzi}}}, \bibinfo {author} {\bibfnamefont {Sajad}\
  \bibnamefont {{Abbar}}}, \bibinfo {author} {\bibfnamefont {Robert}\
  \bibnamefont {{Bollig}}}, \ and\ \bibinfo {author} {\bibfnamefont
  {H.~Thomas}\ \bibnamefont {{Janka}}},\ }\bibfield  {title} {\enquote
  {\bibinfo {title} {{Fast neutrino flavor conversions in one-dimensional
  core-collapse supernova models with and without muon creation}},}\ }\href
  {\doibase 10.1103/PhysRevD.103.063013} {\bibfield  {journal} {\bibinfo
  {journal} {\prd}\ }\textbf {\bibinfo {volume} {103}},\ \bibinfo {eid}
  {063013} (\bibinfo {year} {2021})},\ \Eprint
  {http://arxiv.org/abs/2012.08525} {arXiv:2012.08525 [astro-ph.HE]}
  \BibitemShut {NoStop}%
\bibitem [{\citenamefont {{Harada}}\ and\ \citenamefont
  {{Nagakura}}(2022)}]{2022ApJ...924..109H}%
  \BibitemOpen
  \bibfield  {author} {\bibinfo {author} {\bibfnamefont {Akira}\ \bibnamefont
  {{Harada}}}\ and\ \bibinfo {author} {\bibfnamefont {Hiroki}\ \bibnamefont
  {{Nagakura}}},\ }\bibfield  {title} {\enquote {\bibinfo {title} {{Prospects
  of Fast Flavor Neutrino Conversion in Rotating Core-collapse Supernovae}},}\
  }\href {\doibase 10.3847/1538-4357/ac38a0} {\bibfield  {journal} {\bibinfo
  {journal} {\apj}\ }\textbf {\bibinfo {volume} {924}},\ \bibinfo {eid} {109}
  (\bibinfo {year} {2022})},\ \Eprint {http://arxiv.org/abs/2110.08291}
  {arXiv:2110.08291 [astro-ph.HE]} \BibitemShut {NoStop}%
\bibitem [{\citenamefont {{Xiong}}\ \emph {et~al.}(2023)\citenamefont
  {{Xiong}}, \citenamefont {{Wu}}, \citenamefont {{Mart{\'\i}nez-Pinedo}},
  \citenamefont {{Fischer}}, \citenamefont {{George}}, \citenamefont {{Lin}},\
  and\ \citenamefont {{Johns}}}]{2023PhRvD.107h3016X}%
  \BibitemOpen
  \bibfield  {author} {\bibinfo {author} {\bibfnamefont {Zewei}\ \bibnamefont
  {{Xiong}}}, \bibinfo {author} {\bibfnamefont {Meng-Ru}\ \bibnamefont {{Wu}}},
  \bibinfo {author} {\bibfnamefont {Gabriel}\ \bibnamefont
  {{Mart{\'\i}nez-Pinedo}}}, \bibinfo {author} {\bibfnamefont {Tobias}\
  \bibnamefont {{Fischer}}}, \bibinfo {author} {\bibfnamefont {Manu}\
  \bibnamefont {{George}}}, \bibinfo {author} {\bibfnamefont {Chun-Yu}\
  \bibnamefont {{Lin}}}, \ and\ \bibinfo {author} {\bibfnamefont {Lucas}\
  \bibnamefont {{Johns}}},\ }\bibfield  {title} {\enquote {\bibinfo {title}
  {{Evolution of collisional neutrino flavor instabilities in spherically
  symmetric supernova models}},}\ }\href {\doibase 10.1103/PhysRevD.107.083016}
  {\bibfield  {journal} {\bibinfo  {journal} {\prd}\ }\textbf {\bibinfo
  {volume} {107}},\ \bibinfo {eid} {083016} (\bibinfo {year} {2023})},\ \Eprint
  {http://arxiv.org/abs/2210.08254} {arXiv:2210.08254 [astro-ph.HE]}
  \BibitemShut {NoStop}%
\bibitem [{\citenamefont {{Nagakura}}(2023)}]{2023PhRvL.130u1401N}%
  \BibitemOpen
  \bibfield  {author} {\bibinfo {author} {\bibfnamefont {Hiroki}\ \bibnamefont
  {{Nagakura}}},\ }\bibfield  {title} {\enquote {\bibinfo {title} {{Roles of
  Fast Neutrino-Flavor Conversion on the Neutrino-Heating Mechanism of
  Core-Collapse Supernova}},}\ }\href {\doibase 10.1103/PhysRevLett.130.211401}
  {\bibfield  {journal} {\bibinfo  {journal} {\prl}\ }\textbf {\bibinfo
  {volume} {130}},\ \bibinfo {eid} {211401} (\bibinfo {year} {2023})},\ \Eprint
  {http://arxiv.org/abs/2301.10785} {arXiv:2301.10785 [astro-ph.HE]}
  \BibitemShut {NoStop}%
\bibitem [{\citenamefont {{Nagakura}}\ \emph
  {et~al.}(2021{\natexlab{b}})\citenamefont {{Nagakura}}, \citenamefont
  {{Burrows}}, \citenamefont {{Vartanyan}},\ and\ \citenamefont
  {{Radice}}}]{2021MNRAS.500..696N}%
  \BibitemOpen
  \bibfield  {author} {\bibinfo {author} {\bibfnamefont {Hiroki}\ \bibnamefont
  {{Nagakura}}}, \bibinfo {author} {\bibfnamefont {Adam}\ \bibnamefont
  {{Burrows}}}, \bibinfo {author} {\bibfnamefont {David}\ \bibnamefont
  {{Vartanyan}}}, \ and\ \bibinfo {author} {\bibfnamefont {David}\ \bibnamefont
  {{Radice}}},\ }\bibfield  {title} {\enquote {\bibinfo {title} {{Core-collapse
  supernova neutrino emission and detection informed by state-of-the-art
  three-dimensional numerical models}},}\ }\href {\doibase
  10.1093/mnras/staa2691} {\bibfield  {journal} {\bibinfo  {journal} {\mnras}\
  }\textbf {\bibinfo {volume} {500}},\ \bibinfo {pages} {696--717} (\bibinfo
  {year} {2021}{\natexlab{b}})},\ \Eprint {http://arxiv.org/abs/2007.05000}
  {arXiv:2007.05000 [astro-ph.HE]} \BibitemShut {NoStop}%
\bibitem [{\citenamefont {{Nagakura}}\ \emph
  {et~al.}(2021{\natexlab{c}})\citenamefont {{Nagakura}}, \citenamefont
  {{Burrows}},\ and\ \citenamefont {{Vartanyan}}}]{2021MNRAS.506.1462N}%
  \BibitemOpen
  \bibfield  {author} {\bibinfo {author} {\bibfnamefont {Hiroki}\ \bibnamefont
  {{Nagakura}}}, \bibinfo {author} {\bibfnamefont {Adam}\ \bibnamefont
  {{Burrows}}}, \ and\ \bibinfo {author} {\bibfnamefont {David}\ \bibnamefont
  {{Vartanyan}}},\ }\bibfield  {title} {\enquote {\bibinfo {title} {{Supernova
  neutrino signals based on long-term axisymmetric simulations}},}\ }\href
  {\doibase 10.1093/mnras/stab1785} {\bibfield  {journal} {\bibinfo  {journal}
  {\mnras}\ }\textbf {\bibinfo {volume} {506}},\ \bibinfo {pages} {1462--1479}
  (\bibinfo {year} {2021}{\natexlab{c}})},\ \Eprint
  {http://arxiv.org/abs/2102.11283} {arXiv:2102.11283 [astro-ph.HE]}
  \BibitemShut {NoStop}%
\bibitem [{\citenamefont {{Eggenberger Andersen}}\ \emph
  {et~al.}(2021)\citenamefont {{Eggenberger Andersen}}, \citenamefont {{Zha}},
  \citenamefont {{da Silva Schneider}}, \citenamefont {{Betranhandy}},
  \citenamefont {{Couch}},\ and\ \citenamefont
  {{O'Connor}}}]{2021ApJ...923..201E}%
  \BibitemOpen
  \bibfield  {author} {\bibinfo {author} {\bibfnamefont {Oliver}\ \bibnamefont
  {{Eggenberger Andersen}}}, \bibinfo {author} {\bibfnamefont {Shuai}\
  \bibnamefont {{Zha}}}, \bibinfo {author} {\bibfnamefont {Andr{\'e}}\
  \bibnamefont {{da Silva Schneider}}}, \bibinfo {author} {\bibfnamefont
  {Aurore}\ \bibnamefont {{Betranhandy}}}, \bibinfo {author} {\bibfnamefont
  {Sean~M.}\ \bibnamefont {{Couch}}}, \ and\ \bibinfo {author} {\bibfnamefont
  {Evan~P.}\ \bibnamefont {{O'Connor}}},\ }\bibfield  {title} {\enquote
  {\bibinfo {title} {{Equation-of-state Dependence of Gravitational Waves in
  Core-collapse Supernovae}},}\ }\href {\doibase 10.3847/1538-4357/ac294c}
  {\bibfield  {journal} {\bibinfo  {journal} {\apj}\ }\textbf {\bibinfo
  {volume} {923}},\ \bibinfo {eid} {201} (\bibinfo {year} {2021})},\ \Eprint
  {http://arxiv.org/abs/2106.09734} {arXiv:2106.09734 [astro-ph.HE]}
  \BibitemShut {NoStop}%
\bibitem [{\citenamefont {{Vartanyan}}\ \emph {et~al.}(2023)\citenamefont
  {{Vartanyan}}, \citenamefont {{Burrows}}, \citenamefont {{Wang}},
  \citenamefont {{Coleman}},\ and\ \citenamefont
  {{White}}}]{2023arXiv230207092V}%
  \BibitemOpen
  \bibfield  {author} {\bibinfo {author} {\bibfnamefont {David}\ \bibnamefont
  {{Vartanyan}}}, \bibinfo {author} {\bibfnamefont {Adam}\ \bibnamefont
  {{Burrows}}}, \bibinfo {author} {\bibfnamefont {Tianshu}\ \bibnamefont
  {{Wang}}}, \bibinfo {author} {\bibfnamefont {Matthew S.~B.}\ \bibnamefont
  {{Coleman}}}, \ and\ \bibinfo {author} {\bibfnamefont {Christopher~J.}\
  \bibnamefont {{White}}},\ }\bibfield  {title} {\enquote {\bibinfo {title}
  {{Gravitational-wave signature of core-collapse supernovae}},}\ }\href
  {\doibase 10.1103/PhysRevD.107.103015} {\bibfield  {journal} {\bibinfo
  {journal} {\prd}\ }\textbf {\bibinfo {volume} {107}},\ \bibinfo {eid}
  {103015} (\bibinfo {year} {2023})},\ \Eprint
  {http://arxiv.org/abs/2302.07092} {arXiv:2302.07092 [astro-ph.HE]}
  \BibitemShut {NoStop}%
\bibitem [{\citenamefont {{Skinner}}\ \emph {et~al.}(2019)\citenamefont
  {{Skinner}}, \citenamefont {{Dolence}}, \citenamefont {{Burrows}},
  \citenamefont {{Radice}},\ and\ \citenamefont
  {{Vartanyan}}}]{2019ApJS..241....7S}%
  \BibitemOpen
  \bibfield  {author} {\bibinfo {author} {\bibfnamefont {M.~Aaron}\
  \bibnamefont {{Skinner}}}, \bibinfo {author} {\bibfnamefont {Joshua~C.}\
  \bibnamefont {{Dolence}}}, \bibinfo {author} {\bibfnamefont {Adam}\
  \bibnamefont {{Burrows}}}, \bibinfo {author} {\bibfnamefont {David}\
  \bibnamefont {{Radice}}}, \ and\ \bibinfo {author} {\bibfnamefont {David}\
  \bibnamefont {{Vartanyan}}},\ }\bibfield  {title} {\enquote {\bibinfo {title}
  {{FORNAX: A Flexible Code for Multiphysics Astrophysical Simulations}},}\
  }\href {\doibase 10.3847/1538-4365/ab007f} {\bibfield  {journal} {\bibinfo
  {journal} {The Astrophysical Journal Supplement Series}\ }\textbf {\bibinfo
  {volume} {241}},\ \bibinfo {eid} {7} (\bibinfo {year} {2019})},\ \Eprint
  {http://arxiv.org/abs/1806.07390} {arXiv:1806.07390 [astro-ph.IM]}
  \BibitemShut {NoStop}%
\bibitem [{\citenamefont {{Burrows}}\ \emph {et~al.}(2020)\citenamefont
  {{Burrows}}, \citenamefont {{Radice}}, \citenamefont {{Vartanyan}},
  \citenamefont {{Nagakura}}, \citenamefont {{Skinner}},\ and\ \citenamefont
  {{Dolence}}}]{2020MNRAS.491.2715B}%
  \BibitemOpen
  \bibfield  {author} {\bibinfo {author} {\bibfnamefont {Adam}\ \bibnamefont
  {{Burrows}}}, \bibinfo {author} {\bibfnamefont {David}\ \bibnamefont
  {{Radice}}}, \bibinfo {author} {\bibfnamefont {David}\ \bibnamefont
  {{Vartanyan}}}, \bibinfo {author} {\bibfnamefont {Hiroki}\ \bibnamefont
  {{Nagakura}}}, \bibinfo {author} {\bibfnamefont {M.~Aaron}\ \bibnamefont
  {{Skinner}}}, \ and\ \bibinfo {author} {\bibfnamefont {Joshua~C.}\
  \bibnamefont {{Dolence}}},\ }\bibfield  {title} {\enquote {\bibinfo {title}
  {{The overarching framework of core-collapse supernova explosions as revealed
  by 3D FORNAX simulations}},}\ }\href {\doibase 10.1093/mnras/stz3223}
  {\bibfield  {journal} {\bibinfo  {journal} {\mnras}\ }\textbf {\bibinfo
  {volume} {491}},\ \bibinfo {pages} {2715--2735} (\bibinfo {year} {2020})},\
  \Eprint {http://arxiv.org/abs/1909.04152} {arXiv:1909.04152 [astro-ph.HE]}
  \BibitemShut {NoStop}%
\bibitem [{\citenamefont {{Burrows}}\ and\ \citenamefont
  {{Vartanyan}}(2021)}]{2021Natur.589...29B}%
  \BibitemOpen
  \bibfield  {author} {\bibinfo {author} {\bibfnamefont {A.}~\bibnamefont
  {{Burrows}}}\ and\ \bibinfo {author} {\bibfnamefont {D.}~\bibnamefont
  {{Vartanyan}}},\ }\bibfield  {title} {\enquote {\bibinfo {title}
  {{Core-collapse supernova explosion theory}},}\ }\href {\doibase
  10.1038/s41586-020-03059-w} {\bibfield  {journal} {\bibinfo  {journal}
  {\nat}\ }\textbf {\bibinfo {volume} {589}},\ \bibinfo {pages} {29--39}
  (\bibinfo {year} {2021})},\ \Eprint {http://arxiv.org/abs/2009.14157}
  {arXiv:2009.14157 [astro-ph.SR]} \BibitemShut {NoStop}%
\bibitem [{\citenamefont {{Vaytet}}\ \emph {et~al.}(2011)\citenamefont
  {{Vaytet}}, \citenamefont {{Audit}}, \citenamefont {{Dubroca}},\ and\
  \citenamefont {{Delahaye}}}]{2011JQSRT.112.1323V}%
  \BibitemOpen
  \bibfield  {author} {\bibinfo {author} {\bibfnamefont {N.~M.~H.}\
  \bibnamefont {{Vaytet}}}, \bibinfo {author} {\bibfnamefont {E.}~\bibnamefont
  {{Audit}}}, \bibinfo {author} {\bibfnamefont {B.}~\bibnamefont {{Dubroca}}},
  \ and\ \bibinfo {author} {\bibfnamefont {F.}~\bibnamefont {{Delahaye}}},\
  }\bibfield  {title} {\enquote {\bibinfo {title} {{A numerical model for
  multigroup radiation hydrodynamics}},}\ }\href {\doibase
  10.1016/j.jqsrt.2011.01.027} {\bibfield  {journal} {\bibinfo  {journal}
  {Journal of Quantitative Spectroscopy and Radiative Transfer}\ }\textbf
  {\bibinfo {volume} {112}},\ \bibinfo {pages} {1323--1335} (\bibinfo {year}
  {2011})},\ \Eprint {http://arxiv.org/abs/1101.4955} {arXiv:1101.4955
  [astro-ph.HE]} \BibitemShut {NoStop}%
\bibitem [{\citenamefont {{Steiner}}\ \emph {et~al.}(2013)\citenamefont
  {{Steiner}}, \citenamefont {{Hempel}},\ and\ \citenamefont
  {{Fischer}}}]{2013ApJ...774...17S}%
  \BibitemOpen
  \bibfield  {author} {\bibinfo {author} {\bibfnamefont {A.~W.}\ \bibnamefont
  {{Steiner}}}, \bibinfo {author} {\bibfnamefont {M.}~\bibnamefont {{Hempel}}},
  \ and\ \bibinfo {author} {\bibfnamefont {T.}~\bibnamefont {{Fischer}}},\
  }\bibfield  {title} {\enquote {\bibinfo {title} {{Core-collapse Supernova
  Equations of State Based on Neutron Star Observations}},}\ }\href {\doibase
  10.1088/0004-637X/774/1/17} {\bibfield  {journal} {\bibinfo  {journal}
  {\apj}\ }\textbf {\bibinfo {volume} {774}},\ \bibinfo {eid} {17} (\bibinfo
  {year} {2013})},\ \Eprint {http://arxiv.org/abs/1207.2184} {arXiv:1207.2184
  [astro-ph.SR]} \BibitemShut {NoStop}%
\bibitem [{\citenamefont {{Burrows}}\ \emph {et~al.}(2006)\citenamefont
  {{Burrows}}, \citenamefont {{Reddy}},\ and\ \citenamefont
  {{Thompson}}}]{2006NuPhA.777..356B}%
  \BibitemOpen
  \bibfield  {author} {\bibinfo {author} {\bibfnamefont {A.}~\bibnamefont
  {{Burrows}}}, \bibinfo {author} {\bibfnamefont {S.}~\bibnamefont {{Reddy}}},
  \ and\ \bibinfo {author} {\bibfnamefont {T.~A.}\ \bibnamefont {{Thompson}}},\
  }\bibfield  {title} {\enquote {\bibinfo {title} {{Neutrino opacities in
  nuclear matter}},}\ }\href {\doibase 10.1016/j.nuclphysa.2004.06.012}
  {\bibfield  {journal} {\bibinfo  {journal} {Nuclear Physics A}\ }\textbf
  {\bibinfo {volume} {777}},\ \bibinfo {pages} {356--394} (\bibinfo {year}
  {2006})},\ \Eprint {http://arxiv.org/abs/astro-ph/0404432} {astro-ph/0404432}
  \BibitemShut {NoStop}%
\bibitem [{\citenamefont {{Horowitz}}\ \emph {et~al.}(2017)\citenamefont
  {{Horowitz}}, \citenamefont {{Caballero}}, \citenamefont {{Lin}},
  \citenamefont {{O'Connor}},\ and\ \citenamefont
  {{Schwenk}}}]{2017PhRvC..95b5801H}%
  \BibitemOpen
  \bibfield  {author} {\bibinfo {author} {\bibfnamefont {C.~J.}\ \bibnamefont
  {{Horowitz}}}, \bibinfo {author} {\bibfnamefont {O.~L.}\ \bibnamefont
  {{Caballero}}}, \bibinfo {author} {\bibfnamefont {Z.}~\bibnamefont {{Lin}}},
  \bibinfo {author} {\bibfnamefont {E.}~\bibnamefont {{O'Connor}}}, \ and\
  \bibinfo {author} {\bibfnamefont {A.}~\bibnamefont {{Schwenk}}},\ }\bibfield
  {title} {\enquote {\bibinfo {title} {{Neutrino-nucleon scattering in
  supernova matter from the virial expansion}},}\ }\href {\doibase
  10.1103/PhysRevC.95.025801} {\bibfield  {journal} {\bibinfo  {journal}
  {\prc}\ }\textbf {\bibinfo {volume} {95}},\ \bibinfo {eid} {025801} (\bibinfo
  {year} {2017})},\ \Eprint {http://arxiv.org/abs/1611.05140} {arXiv:1611.05140
  [nucl-th]} \BibitemShut {NoStop}%
\bibitem [{\citenamefont {{Juodagalvis}}\ \emph {et~al.}(2010)\citenamefont
  {{Juodagalvis}}, \citenamefont {{Langanke}}, \citenamefont {{Hix}},
  \citenamefont {{Mart{\'{\i}}nez-Pinedo}},\ and\ \citenamefont
  {{Sampaio}}}]{2010NuPhA.848..454J}%
  \BibitemOpen
  \bibfield  {author} {\bibinfo {author} {\bibfnamefont {A.}~\bibnamefont
  {{Juodagalvis}}}, \bibinfo {author} {\bibfnamefont {K.}~\bibnamefont
  {{Langanke}}}, \bibinfo {author} {\bibfnamefont {W.~R.}\ \bibnamefont
  {{Hix}}}, \bibinfo {author} {\bibfnamefont {G.}~\bibnamefont
  {{Mart{\'{\i}}nez-Pinedo}}}, \ and\ \bibinfo {author} {\bibfnamefont {J.~M.}\
  \bibnamefont {{Sampaio}}},\ }\bibfield  {title} {\enquote {\bibinfo {title}
  {{Improved estimate of electron capture rates on nuclei during stellar core
  collapse}},}\ }\href {\doibase 10.1016/j.nuclphysa.2010.09.012} {\bibfield
  {journal} {\bibinfo  {journal} {Nuclear Physics A}\ }\textbf {\bibinfo
  {volume} {848}},\ \bibinfo {pages} {454--478} (\bibinfo {year} {2010})},\
  \Eprint {http://arxiv.org/abs/0909.0179} {arXiv:0909.0179 [nucl-th]}
  \BibitemShut {NoStop}%
\bibitem [{\citenamefont {{Vartanyan}}\ and\ \citenamefont
  {{Burrows}}(2023)}]{2023arXiv230708735V}%
  \BibitemOpen
  \bibfield  {author} {\bibinfo {author} {\bibfnamefont {David}\ \bibnamefont
  {{Vartanyan}}}\ and\ \bibinfo {author} {\bibfnamefont {Adam}\ \bibnamefont
  {{Burrows}}},\ }\bibfield  {title} {\enquote {\bibinfo {title} {{Neutrino
  Signatures of One Hundred 2D Axisymmetric Core-Collapse Supernova
  Simulations}},}\ }\href {\doibase 10.48550/arXiv.2307.08735} {\bibfield
  {journal} {\bibinfo  {journal} {arXiv e-prints}\ ,\ \bibinfo {eid}
  {arXiv:2307.08735}} (\bibinfo {year} {2023})},\ \Eprint
  {http://arxiv.org/abs/2307.08735} {arXiv:2307.08735 [astro-ph.HE]}
  \BibitemShut {NoStop}%
\bibitem [{\citenamefont {{Abe}}\ and\ \citenamefont
  {et~al.}(2016)}]{2016APh....81...39A}%
  \BibitemOpen
  \bibfield  {author} {\bibinfo {author} {\bibfnamefont {K.}~\bibnamefont
  {{Abe}}}\ and\ \bibinfo {author} {\bibnamefont {et~al.}},\ }\bibfield
  {title} {\enquote {\bibinfo {title} {{Real-time supernova neutrino burst
  monitor at Super-Kamiokande}},}\ }\href {\doibase
  10.1016/j.astropartphys.2016.04.003} {\bibfield  {journal} {\bibinfo
  {journal} {Astroparticle Physics}\ }\textbf {\bibinfo {volume} {81}},\
  \bibinfo {pages} {39--48} (\bibinfo {year} {2016})},\ \Eprint
  {http://arxiv.org/abs/1601.04778} {arXiv:1601.04778 [astro-ph.HE]}
  \BibitemShut {NoStop}%
\bibitem [{\citenamefont {{An}}\ \emph {et~al.}(2016)\citenamefont {{An}},
  \citenamefont {{An}}, \citenamefont {{An}}, \citenamefont {{Antonelli}},
  \citenamefont {{Baussan}}, \citenamefont {{Beacom}}, \citenamefont
  {{Bezrukov}}, \citenamefont {{Blyth}}, \citenamefont {{Brugnera}},
  \citenamefont {{Buizza Avanzini}},\ and\ \citenamefont
  {et~al.}}]{2016JPhG...43c0401A}%
  \BibitemOpen
  \bibfield  {author} {\bibinfo {author} {\bibfnamefont {Fengpeng}\
  \bibnamefont {{An}}}, \bibinfo {author} {\bibfnamefont {Guangpeng}\
  \bibnamefont {{An}}}, \bibinfo {author} {\bibfnamefont {Qi}~\bibnamefont
  {{An}}}, \bibinfo {author} {\bibfnamefont {Vito}\ \bibnamefont
  {{Antonelli}}}, \bibinfo {author} {\bibfnamefont {Eric}\ \bibnamefont
  {{Baussan}}}, \bibinfo {author} {\bibfnamefont {John}\ \bibnamefont
  {{Beacom}}}, \bibinfo {author} {\bibfnamefont {Leonid}\ \bibnamefont
  {{Bezrukov}}}, \bibinfo {author} {\bibfnamefont {Simon}\ \bibnamefont
  {{Blyth}}}, \bibinfo {author} {\bibfnamefont {Riccardo}\ \bibnamefont
  {{Brugnera}}}, \bibinfo {author} {\bibfnamefont {Margherita}\ \bibnamefont
  {{Buizza Avanzini}}}, \ and\ \bibinfo {author} {\bibnamefont {et~al.}},\
  }\bibfield  {title} {\enquote {\bibinfo {title} {{Neutrino physics with
  JUNO}},}\ }\href {\doibase 10.1088/0954-3899/43/3/030401} {\bibfield
  {journal} {\bibinfo  {journal} {Journal of Physics G Nuclear Physics}\
  }\textbf {\bibinfo {volume} {43}},\ \bibinfo {eid} {030401} (\bibinfo {year}
  {2016})},\ \Eprint {http://arxiv.org/abs/1507.05613} {arXiv:1507.05613
  [physics.ins-det]} \BibitemShut {NoStop}%
\bibitem [{\citenamefont {{Nagakura}}\ and\ \citenamefont
  {{Vartanyan}}(2022)}]{2022MNRAS.512.2806N}%
  \BibitemOpen
  \bibfield  {author} {\bibinfo {author} {\bibfnamefont {Hiroki}\ \bibnamefont
  {{Nagakura}}}\ and\ \bibinfo {author} {\bibfnamefont {David}\ \bibnamefont
  {{Vartanyan}}},\ }\bibfield  {title} {\enquote {\bibinfo {title} {{Efficient
  method for estimating the time evolution of the proto-neutron star mass and
  radius from a supernova neutrino signal}},}\ }\href {\doibase
  10.1093/mnras/stac383} {\bibfield  {journal} {\bibinfo  {journal} {\mnras}\
  }\textbf {\bibinfo {volume} {512}},\ \bibinfo {pages} {2806--2816} (\bibinfo
  {year} {2022})},\ \Eprint {http://arxiv.org/abs/2111.05869} {arXiv:2111.05869
  [astro-ph.HE]} \BibitemShut {NoStop}%
\bibitem [{\citenamefont {{Halzen}}\ and\ \citenamefont
  {{Raffelt}}(2009)}]{2009PhRvD..80h7301H}%
  \BibitemOpen
  \bibfield  {author} {\bibinfo {author} {\bibfnamefont {Francis}\ \bibnamefont
  {{Halzen}}}\ and\ \bibinfo {author} {\bibfnamefont {Georg~G.}\ \bibnamefont
  {{Raffelt}}},\ }\bibfield  {title} {\enquote {\bibinfo {title}
  {{Reconstructing the supernova bounce time with neutrinos in IceCube}},}\
  }\href {\doibase 10.1103/PhysRevD.80.087301} {\bibfield  {journal} {\bibinfo
  {journal} {\prd}\ }\textbf {\bibinfo {volume} {80}},\ \bibinfo {eid} {087301}
  (\bibinfo {year} {2009})},\ \Eprint {http://arxiv.org/abs/0908.2317}
  {arXiv:0908.2317 [astro-ph.HE]} \BibitemShut {NoStop}%
\bibitem [{\citenamefont {{Vartanyan}}\ \emph {et~al.}(2019)\citenamefont
  {{Vartanyan}}, \citenamefont {{Burrows}},\ and\ \citenamefont
  {{Radice}}}]{2019MNRAS.489.2227V}%
  \BibitemOpen
  \bibfield  {author} {\bibinfo {author} {\bibfnamefont {David}\ \bibnamefont
  {{Vartanyan}}}, \bibinfo {author} {\bibfnamefont {Adam}\ \bibnamefont
  {{Burrows}}}, \ and\ \bibinfo {author} {\bibfnamefont {David}\ \bibnamefont
  {{Radice}}},\ }\bibfield  {title} {\enquote {\bibinfo {title} {{Temporal and
  angular variations of 3D core-collapse supernova emissions and their physical
  correlations}},}\ }\href {\doibase 10.1093/mnras/stz2307} {\bibfield
  {journal} {\bibinfo  {journal} {\mnras}\ }\textbf {\bibinfo {volume} {489}},\
  \bibinfo {pages} {2227--2246} (\bibinfo {year} {2019})},\ \Eprint
  {http://arxiv.org/abs/1906.08787} {arXiv:1906.08787 [astro-ph.HE]}
  \BibitemShut {NoStop}%
\bibitem [{\citenamefont {{Vartanyan}}\ and\ \citenamefont
  {{Burrows}}(2020)}]{2020ApJ...901..108V}%
  \BibitemOpen
  \bibfield  {author} {\bibinfo {author} {\bibfnamefont {David}\ \bibnamefont
  {{Vartanyan}}}\ and\ \bibinfo {author} {\bibfnamefont {Adam}\ \bibnamefont
  {{Burrows}}},\ }\bibfield  {title} {\enquote {\bibinfo {title}
  {{Gravitational Waves from Neutrino Emission Asymmetries in Core-collapse
  Supernovae}},}\ }\href {\doibase 10.3847/1538-4357/abafac} {\bibfield
  {journal} {\bibinfo  {journal} {\apj}\ }\textbf {\bibinfo {volume} {901}},\
  \bibinfo {eid} {108} (\bibinfo {year} {2020})},\ \Eprint
  {http://arxiv.org/abs/2007.07261} {arXiv:2007.07261 [astro-ph.HE]}
  \BibitemShut {NoStop}%
\bibitem [{\citenamefont {{Mueller}}\ and\ \citenamefont
  {{Janka}}(1997)}]{1997A&A...317..140M}%
  \BibitemOpen
  \bibfield  {author} {\bibinfo {author} {\bibfnamefont {E.}~\bibnamefont
  {{Mueller}}}\ and\ \bibinfo {author} {\bibfnamefont {H.~T.}\ \bibnamefont
  {{Janka}}},\ }\bibfield  {title} {\enquote {\bibinfo {title} {{Gravitational
  radiation from convective instabilities in Type II supernova explosions.}}}\
  }\href@noop {} {\bibfield  {journal} {\bibinfo  {journal} {\aap}\ }\textbf
  {\bibinfo {volume} {317}},\ \bibinfo {pages} {140--163} (\bibinfo {year}
  {1997})}\BibitemShut {NoStop}%
\bibitem [{\citenamefont {{Wang}}\ \emph {et~al.}(2022)\citenamefont {{Wang}},
  \citenamefont {{Vartanyan}}, \citenamefont {{Burrows}},\ and\ \citenamefont
  {{Coleman}}}]{2022MNRAS.517..543W}%
  \BibitemOpen
  \bibfield  {author} {\bibinfo {author} {\bibfnamefont {Tianshu}\ \bibnamefont
  {{Wang}}}, \bibinfo {author} {\bibfnamefont {David}\ \bibnamefont
  {{Vartanyan}}}, \bibinfo {author} {\bibfnamefont {Adam}\ \bibnamefont
  {{Burrows}}}, \ and\ \bibinfo {author} {\bibfnamefont {Matthew S.~B.}\
  \bibnamefont {{Coleman}}},\ }\bibfield  {title} {\enquote {\bibinfo {title}
  {{The essential character of the neutrino mechanism of core-collapse
  supernova explosions}},}\ }\href {\doibase 10.1093/mnras/stac2691} {\bibfield
   {journal} {\bibinfo  {journal} {\mnras}\ }\textbf {\bibinfo {volume}
  {517}},\ \bibinfo {pages} {543--559} (\bibinfo {year} {2022})},\ \Eprint
  {http://arxiv.org/abs/2207.02231} {arXiv:2207.02231 [astro-ph.SR]}
  \BibitemShut {NoStop}%
\bibitem [{\citenamefont {{Ugliano}}\ \emph {et~al.}(2012)\citenamefont
  {{Ugliano}}, \citenamefont {{Janka}}, \citenamefont {{Marek}},\ and\
  \citenamefont {{Arcones}}}]{2012ApJ...757...69U}%
  \BibitemOpen
  \bibfield  {author} {\bibinfo {author} {\bibfnamefont {Marcella}\
  \bibnamefont {{Ugliano}}}, \bibinfo {author} {\bibfnamefont {Hans-Thomas}\
  \bibnamefont {{Janka}}}, \bibinfo {author} {\bibfnamefont {Andreas}\
  \bibnamefont {{Marek}}}, \ and\ \bibinfo {author} {\bibfnamefont {Almudena}\
  \bibnamefont {{Arcones}}},\ }\bibfield  {title} {\enquote {\bibinfo {title}
  {{Progenitor-explosion Connection and Remnant Birth Masses for
  Neutrino-driven Supernovae of Iron-core Progenitors}},}\ }\href {\doibase
  10.1088/0004-637X/757/1/69} {\bibfield  {journal} {\bibinfo  {journal}
  {\apj}\ }\textbf {\bibinfo {volume} {757}},\ \bibinfo {eid} {69} (\bibinfo
  {year} {2012})},\ \Eprint {http://arxiv.org/abs/1205.3657} {arXiv:1205.3657
  [astro-ph.SR]} \BibitemShut {NoStop}%
\bibitem [{\citenamefont {{Finn}}\ and\ \citenamefont
  {{Evans}}(1990)}]{1990ApJ...351..588F}%
  \BibitemOpen
  \bibfield  {author} {\bibinfo {author} {\bibfnamefont {Lee~Samuel}\
  \bibnamefont {{Finn}}}\ and\ \bibinfo {author} {\bibfnamefont {Charles~R.}\
  \bibnamefont {{Evans}}},\ }\bibfield  {title} {\enquote {\bibinfo {title}
  {{Determing gravitational radiation from Newtonian self-gravitating
  systems.}}}\ }\href {\doibase 10.1086/168497} {\bibfield  {journal} {\bibinfo
   {journal} {\apj}\ }\textbf {\bibinfo {volume} {351}},\ \bibinfo {pages}
  {588--600} (\bibinfo {year} {1990})}\BibitemShut {NoStop}%
\bibitem [{\citenamefont {{Nagakura}}(2021)}]{2021MNRAS.500..319N}%
  \BibitemOpen
  \bibfield  {author} {\bibinfo {author} {\bibfnamefont {Hiroki}\ \bibnamefont
  {{Nagakura}}},\ }\bibfield  {title} {\enquote {\bibinfo {title} {{Retrieval
  of energy spectra for all flavours of neutrinos from core-collapse supernova
  with multiple detectors}},}\ }\href {\doibase 10.1093/mnras/staa3287}
  {\bibfield  {journal} {\bibinfo  {journal} {\mnras}\ }\textbf {\bibinfo
  {volume} {500}},\ \bibinfo {pages} {319--332} (\bibinfo {year} {2021})},\
  \Eprint {http://arxiv.org/abs/2008.10082} {arXiv:2008.10082 [astro-ph.HE]}
  \BibitemShut {NoStop}%
\bibitem [{\citenamefont {{Beacom}}\ and\ \citenamefont
  {{Vagins}}(2004)}]{2004PhRvL..93q1101B}%
  \BibitemOpen
  \bibfield  {author} {\bibinfo {author} {\bibfnamefont {John~F.}\ \bibnamefont
  {{Beacom}}}\ and\ \bibinfo {author} {\bibfnamefont {Mark~R.}\ \bibnamefont
  {{Vagins}}},\ }\bibfield  {title} {\enquote {\bibinfo {title} {{Antineutrino
  Spectroscopy with Large Water {\v{C}}erenkov Detectors}},}\ }\href {\doibase
  10.1103/PhysRevLett.93.171101} {\bibfield  {journal} {\bibinfo  {journal}
  {\prl}\ }\textbf {\bibinfo {volume} {93}},\ \bibinfo {eid} {171101} (\bibinfo
  {year} {2004})},\ \Eprint {http://arxiv.org/abs/hep-ph/0309300}
  {arXiv:hep-ph/0309300 [hep-ph]} \BibitemShut {NoStop}%
\bibitem [{\citenamefont {{Harada}}(2023)}]{2023arXiv230505135H}%
  \BibitemOpen
  \bibfield  {author} {\bibinfo {author} {\bibfnamefont {M.~et~al.}\
  \bibnamefont {{Harada}}},\ }\bibfield  {title} {\enquote {\bibinfo {title}
  {{Search for astrophysical electron antineutrinos in Super-Kamiokande with
  0.01wt\% gadolinium-loaded water}},}\ }\href {\doibase
  10.48550/arXiv.2305.05135} {\bibfield  {journal} {\bibinfo  {journal} {arXiv
  e-prints}\ ,\ \bibinfo {eid} {arXiv:2305.05135}} (\bibinfo {year} {2023})},\
  \Eprint {http://arxiv.org/abs/2305.05135} {arXiv:2305.05135 [astro-ph.HE]}
  \BibitemShut {NoStop}%
\bibitem [{\citenamefont {{Hosokawa}}\ \emph {et~al.}(2023)\citenamefont
  {{Hosokawa}}, \citenamefont {{Ikeda}}, \citenamefont {{Okada}}, \citenamefont
  {{Sekiya}}, \citenamefont {{Fern{\'a}ndez}}, \citenamefont {{Labarga}},
  \citenamefont {{Bandac}}, \citenamefont {{Perez}}, \citenamefont {{Ito}},
  \citenamefont {{Harada}}, \citenamefont {{Koshio}}, \citenamefont
  {{Thiesse}}, \citenamefont {{Thompson}}, \citenamefont {{Scovell}},
  \citenamefont {{Meehan}}, \citenamefont {{Ichimura}}, \citenamefont
  {{Kishimoto}}, \citenamefont {{Nakajima}}, \citenamefont {{Vagins}},
  \citenamefont {{Ito}}, \citenamefont {{Takaku}}, \citenamefont {{Tanaka}},\
  and\ \citenamefont {{Yamaguchi}}}]{2023PTEP.2023a3H01H}%
  \BibitemOpen
  \bibfield  {author} {\bibinfo {author} {\bibfnamefont {K.}~\bibnamefont
  {{Hosokawa}}}, \bibinfo {author} {\bibfnamefont {M.}~\bibnamefont {{Ikeda}}},
  \bibinfo {author} {\bibfnamefont {T.}~\bibnamefont {{Okada}}}, \bibinfo
  {author} {\bibfnamefont {H.}~\bibnamefont {{Sekiya}}}, \bibinfo {author}
  {\bibfnamefont {P.}~\bibnamefont {{Fern{\'a}ndez}}}, \bibinfo {author}
  {\bibfnamefont {L.}~\bibnamefont {{Labarga}}}, \bibinfo {author}
  {\bibfnamefont {I.}~\bibnamefont {{Bandac}}}, \bibinfo {author}
  {\bibfnamefont {J.}~\bibnamefont {{Perez}}}, \bibinfo {author} {\bibfnamefont
  {S.}~\bibnamefont {{Ito}}}, \bibinfo {author} {\bibfnamefont
  {M.}~\bibnamefont {{Harada}}}, \bibinfo {author} {\bibfnamefont
  {Y.}~\bibnamefont {{Koshio}}}, \bibinfo {author} {\bibfnamefont {M.~D.}\
  \bibnamefont {{Thiesse}}}, \bibinfo {author} {\bibfnamefont {L.~F.}\
  \bibnamefont {{Thompson}}}, \bibinfo {author} {\bibfnamefont {P.~R.}\
  \bibnamefont {{Scovell}}}, \bibinfo {author} {\bibfnamefont {E.}~\bibnamefont
  {{Meehan}}}, \bibinfo {author} {\bibfnamefont {K.}~\bibnamefont
  {{Ichimura}}}, \bibinfo {author} {\bibfnamefont {Y.}~\bibnamefont
  {{Kishimoto}}}, \bibinfo {author} {\bibfnamefont {Y.}~\bibnamefont
  {{Nakajima}}}, \bibinfo {author} {\bibfnamefont {M.~R.}\ \bibnamefont
  {{Vagins}}}, \bibinfo {author} {\bibfnamefont {H.}~\bibnamefont {{Ito}}},
  \bibinfo {author} {\bibfnamefont {Y.}~\bibnamefont {{Takaku}}}, \bibinfo
  {author} {\bibfnamefont {Y.}~\bibnamefont {{Tanaka}}}, \ and\ \bibinfo
  {author} {\bibfnamefont {Y.}~\bibnamefont {{Yamaguchi}}},\ }\bibfield
  {title} {\enquote {\bibinfo {title} {{Development of ultra-pure gadolinium
  sulfate for the Super-Kamiokande gadolinium project}},}\ }\href {\doibase
  10.1093/ptep/ptac170} {\bibfield  {journal} {\bibinfo  {journal} {Progress of
  Theoretical and Experimental Physics}\ }\textbf {\bibinfo {volume} {2023}},\
  \bibinfo {eid} {013H01} (\bibinfo {year} {2023})},\ \Eprint
  {http://arxiv.org/abs/2209.07273} {arXiv:2209.07273 [physics.ins-det]}
  \BibitemShut {NoStop}%
\bibitem [{\citenamefont {{Braginskii}}\ and\ \citenamefont
  {{Thorne}}(1987)}]{1987Natur.327..123B}%
  \BibitemOpen
  \bibfield  {author} {\bibinfo {author} {\bibfnamefont {V.~B.}\ \bibnamefont
  {{Braginskii}}}\ and\ \bibinfo {author} {\bibfnamefont {Kip~S.}\ \bibnamefont
  {{Thorne}}},\ }\bibfield  {title} {\enquote {\bibinfo {title}
  {{Gravitational-wave bursts with memory and experimental prospects}},}\
  }\href {\doibase 10.1038/327123a0} {\bibfield  {journal} {\bibinfo  {journal}
  {\nat}\ }\textbf {\bibinfo {volume} {327}},\ \bibinfo {pages} {123--125}
  (\bibinfo {year} {1987})}\BibitemShut {NoStop}%
\bibitem [{\citenamefont {{Kotake}}\ \emph {et~al.}(2007)\citenamefont
  {{Kotake}}, \citenamefont {{Ohnishi}},\ and\ \citenamefont
  {{Yamada}}}]{2007ApJ...655..406K}%
  \BibitemOpen
  \bibfield  {author} {\bibinfo {author} {\bibfnamefont {Kei}\ \bibnamefont
  {{Kotake}}}, \bibinfo {author} {\bibfnamefont {Naofumi}\ \bibnamefont
  {{Ohnishi}}}, \ and\ \bibinfo {author} {\bibfnamefont {Shoichi}\ \bibnamefont
  {{Yamada}}},\ }\bibfield  {title} {\enquote {\bibinfo {title} {{Gravitational
  Radiation from Standing Accretion Shock Instability in Core-Collapse
  Supernovae}},}\ }\href {\doibase 10.1086/509320} {\bibfield  {journal}
  {\bibinfo  {journal} {\apj}\ }\textbf {\bibinfo {volume} {655}},\ \bibinfo
  {pages} {406--415} (\bibinfo {year} {2007})},\ \Eprint
  {http://arxiv.org/abs/astro-ph/0607224} {arXiv:astro-ph/0607224 [astro-ph]}
  \BibitemShut {NoStop}%
\bibitem [{\citenamefont {{Kotake}}\ \emph {et~al.}(2009)\citenamefont
  {{Kotake}}, \citenamefont {{Iwakami}}, \citenamefont {{Ohnishi}},\ and\
  \citenamefont {{Yamada}}}]{2009ApJ...697L.133K}%
  \BibitemOpen
  \bibfield  {author} {\bibinfo {author} {\bibfnamefont {Kei}\ \bibnamefont
  {{Kotake}}}, \bibinfo {author} {\bibfnamefont {Wakana}\ \bibnamefont
  {{Iwakami}}}, \bibinfo {author} {\bibfnamefont {Naofumi}\ \bibnamefont
  {{Ohnishi}}}, \ and\ \bibinfo {author} {\bibfnamefont {Shoichi}\ \bibnamefont
  {{Yamada}}},\ }\bibfield  {title} {\enquote {\bibinfo {title} {{Stochastic
  Nature of Gravitational Waves from Supernova Explosions with Standing
  Accretion Shock Instability}},}\ }\href {\doibase
  10.1088/0004-637X/697/2/L133} {\bibfield  {journal} {\bibinfo  {journal}
  {\apjl}\ }\textbf {\bibinfo {volume} {697}},\ \bibinfo {pages} {L133--L136}
  (\bibinfo {year} {2009})},\ \Eprint {http://arxiv.org/abs/0904.4300}
  {arXiv:0904.4300 [astro-ph.HE]} \BibitemShut {NoStop}%
\bibitem [{\citenamefont {{M{\"u}ller}}\ \emph {et~al.}(2012)\citenamefont
  {{M{\"u}ller}}, \citenamefont {{Janka}},\ and\ \citenamefont
  {{Wongwathanarat}}}]{2012A&A...537A..63M}%
  \BibitemOpen
  \bibfield  {author} {\bibinfo {author} {\bibfnamefont {E.}~\bibnamefont
  {{M{\"u}ller}}}, \bibinfo {author} {\bibfnamefont {H.~Th.}\ \bibnamefont
  {{Janka}}}, \ and\ \bibinfo {author} {\bibfnamefont {A.}~\bibnamefont
  {{Wongwathanarat}}},\ }\bibfield  {title} {\enquote {\bibinfo {title}
  {{Parametrized 3D models of neutrino-driven supernova explosions. Neutrino
  emission asymmetries and gravitational-wave signals}},}\ }\href {\doibase
  10.1051/0004-6361/201117611} {\bibfield  {journal} {\bibinfo  {journal}
  {\aap}\ }\textbf {\bibinfo {volume} {537}},\ \bibinfo {eid} {A63} (\bibinfo
  {year} {2012})},\ \Eprint {http://arxiv.org/abs/1106.6301} {arXiv:1106.6301
  [astro-ph.SR]} \BibitemShut {NoStop}%
\bibitem [{\citenamefont {{Fu}}\ and\ \citenamefont
  {{Yamada}}(2022)}]{2022PhRvD.105l3028F}%
  \BibitemOpen
  \bibfield  {author} {\bibinfo {author} {\bibfnamefont {Lei}\ \bibnamefont
  {{Fu}}}\ and\ \bibinfo {author} {\bibfnamefont {Shoichi}\ \bibnamefont
  {{Yamada}}},\ }\bibfield  {title} {\enquote {\bibinfo {title} {{Gravitational
  wave signals in the deci-Hz range from neutrinos during the protoneutron star
  cooling phase}},}\ }\href {\doibase 10.1103/PhysRevD.105.123028} {\bibfield
  {journal} {\bibinfo  {journal} {\prd}\ }\textbf {\bibinfo {volume} {105}},\
  \bibinfo {eid} {123028} (\bibinfo {year} {2022})},\ \Eprint
  {http://arxiv.org/abs/2201.12774} {arXiv:2201.12774 [astro-ph.HE]}
  \BibitemShut {NoStop}%
\bibitem [{\citenamefont {{Kachelrie{\ss}}}\ \emph {et~al.}(2005)\citenamefont
  {{Kachelrie{\ss}}}, \citenamefont {{Tom{\`a}s}}, \citenamefont {{Buras}},
  \citenamefont {{Janka}}, \citenamefont {{Marek}},\ and\ \citenamefont
  {{Rampp}}}]{2005PhRvD..71f3003K}%
  \BibitemOpen
  \bibfield  {author} {\bibinfo {author} {\bibfnamefont {M.}~\bibnamefont
  {{Kachelrie{\ss}}}}, \bibinfo {author} {\bibfnamefont {R.}~\bibnamefont
  {{Tom{\`a}s}}}, \bibinfo {author} {\bibfnamefont {R.}~\bibnamefont
  {{Buras}}}, \bibinfo {author} {\bibfnamefont {H.-T.}\ \bibnamefont
  {{Janka}}}, \bibinfo {author} {\bibfnamefont {A.}~\bibnamefont {{Marek}}}, \
  and\ \bibinfo {author} {\bibfnamefont {M.}~\bibnamefont {{Rampp}}},\
  }\bibfield  {title} {\enquote {\bibinfo {title} {{Exploiting the
  neutronization burst of a galactic supernova}},}\ }\href {\doibase
  10.1103/PhysRevD.71.063003} {\bibfield  {journal} {\bibinfo  {journal}
  {\prd}\ }\textbf {\bibinfo {volume} {71}},\ \bibinfo {eid} {063003} (\bibinfo
  {year} {2005})},\ \Eprint {http://arxiv.org/abs/astro-ph/0412082}
  {astro-ph/0412082} \BibitemShut {NoStop}%
\end{thebibliography}%

\end{document}